\documentclass[prx,twocolumn,groupedaddress]{revtex4-2}
\usepackage{amsmath,amsfonts,amssymb,graphicx,xcolor,ulem}

\newcommand{\ve}[1]{\mathbf{#1}}

\begin{document}

%\preprint{}

\title{Temporal coarse-graining and elimination of slow dynamics with the  generalized Langevin equation for time-filtered observables }


\author{Roland R. Netz}
%\thanks{}
\affiliation{Fachbereich Physik, Freie Universit\"at Berlin, 144195 Berlin, Germany}
\affiliation{Centre for Condensed Matter Theory, Department of Physics, Indian Institute of Science, Bangalore 560012, India}


\date{\today}

\begin{abstract}
% insert abstract here: PRX < 500 words
By exact projection in phase space we  derive the  generalized Langevin equation (GLE) 
 for  time-filtered  observables. We employ a general convolution filter that
  directly acts on arbitrary phase-space observables and can involve
   low-pass, high-pass, band-pass or band-stop components.
  The derived filter GLE has the same form and properties as the ordinary GLE but exhibits
modified  potential,  mass and  memory friction kernel.  Our filter-projection approach has   diverse 
applications and   can be used to
  i)  systematically derive temporally coarse-grained models by low-pass filtering,
  ii) undo data smoothening  inherent in any experimental measurement process,
  iii) decompose data exactly into slow and fast variables that can be analyzed separately and each obey Liouville  dynamics.
  The latter application is suitable for removing slow transient  or seasonal (i.e. periodic)  components  that do not
  equilibrate over  simulation or experimental observation time scales
and constitutes an alternative  to  non-equilibrium approaches.
  We derive  integral formulas for the  GLE parameters  of filtered data for general systems.
  For the special case of a Markovian system we derive the filter GLE memory kernel  in closed form and
  show that low-band pass  smoothening of data  induces exponentially decaying memory.
  \end{abstract}

%\keywords{Statistical Physics, Non-Markovian Effects, Spatially Dependent Memory}
\maketitle
%Popular Summary: Physical Review X requires authors to submit a nontechnical summary that conveys the context, the essential message(s), 
%and the significance of the work to all readers. The summary should be concise (approximately 250 words), readable, objective, and have broad appeal. 
%Please avoid including mathematical expressions.
\section{Introduction}

Doing science often means  understanding time-series data or trajectories, which are measurements  of observables  over time
and  stem from computer simulations, at varying levels of modeling, or from experiments. 
When analyzing such  trajectories, the problem of   time-scale multiplicity  arises:
In many cases,
the process of interest is coupled to microscopic dynamics at much shorter time scales, 
e.g., observables describing  protein-folding dynamics at  milliseconds  to seconds often include contributions from molecular vibrations
 on the femtosecond time scale and from water-binding dynamics at the picosecond time scale 
 \cite{Levitt1975,Kmiecik2016}.
 In other cases,  trajectories are   perturbed by processes taking place on much longer times than one is interested in.
 For example, motility experiments on living cells are influenced by cell-division dynamics \cite{walmod2004}, 
 daily weather dynamics is coupled to  seasonal weather changes \cite{franzke2015stochastic}.
 The concept of time-scale separation describes idealized situations where time scales much shorter or longer 
 than the time scale of interest need not be considered since they  essentially decouple \cite{Gunawardena2013},
 here we are interested in systems where this decoupling does not take place. 
 
The problem of short and  in many applications irrelevant time scales in experimental or simulated  trajectories is often 
simply ignored by recording data at a sufficiently low resolution, so that fast processes are not captured. 
This approach is unsatisfactory,  since it is difficult to construct a model that  captures all aspects of 
 the long-time system dynamics from highly discretized  data \cite{Tepper2024}.
A more systematic approach to eliminate fast processes is coarse-graining, which allows to build molecular
models that do only    include a subset of  essential  microscopic degrees of freedom. An early example are
 molecular dynamics models that replace  electronic degrees of freedom by effective
atomistic  interaction potentials, current coarse-grained models  unite atoms  into molecular groups
and  molecules into supramolecules and are used to describe large-scale protein, RNA and DNA properties 
\cite{Levitt1975,Thirumalai2011,Thirumalai2013,Doye2013,Papoian2013,Kmiecik2016,Marrink2021}.
By reducing the number of degrees of freedom, the resulting coarse-grained models
not only require fewer  computations per time step but also eliminate fast dynamics, so that the 
simulation time step can be increased. The challenge of coarse graining is
to define suitable degrees of freedom and to construct their effective interactions, 
which can be  met by machine-learning techniques \cite{Clementi2019}. 
However, an essentially unsolved problem is that the coarse-grained dynamic equation of motion
becomes non-Markovian, which is relevant  for recovering  correct long-time dynamics but  is difficult
to derive and to  implement numerically \cite{ayaz2021,dalton2023}. 
The   filtering-projection approach  introduced in this paper  can be used to systematically construct 
temporally coarse-grained models  with systematically reduced fast dynamics 
that  accurately  preserve   the long-time observable dynamics.

The converse problem of  long time scales that are present in the system and  that couple to the process of interest  is  even more subtle. 
It is undisputed that the Hamiltonian of a large enough system, which eventually would encompass
the entire universe,  becomes autonomous and thus independent of time \cite{schmitt_analyzing_2006}, in which limit  the standard time-independent 
Liouville operator describes the time evolution of the system and in the stationary state (if it exists and is reached
on  relevant time scales) equilibrium statistical mechanics applies.  This changes when the described system is made finite
and instead put in contact with a time-varying environment, in which case the Hamiltonian becomes time-dependent.
Here one deals with an intrinsically non-equilibrium situation, i.e. the system is typically far from equilibrium even 
in its stationary state (if there is one) and the needed theoretical concepts are much less developed than for equilibrium scenarios
\cite{Mazur1953,Lebowitz1959,zwanzig_ensemble_1960,deGroot,Zia,Derrida2001,Barrat2007,Fodor2016,
 Jarzynski2000,Harada2005,Seifert2005,Prost2009,Wynants2009,Seifert2010,Netz2020}.
   Also for these situations, our  filtering-projection approach   can be used to systematically eliminate
the long-time  transient or oscillatory data components and thus circumvents the need to deal with non-equilibrium effects in the data.

The equation that exactly describes the dynamics of an arbitrary observable over time,
for example the position or shape of a molecule,
the degree of folding  of a protein or the reaction coordinate describing a chemical reaction, 
 is the generalized Langevin equation (GLE) \cite{mori_transport_1965,zwanzig1961memory}, 
which is an  integro-differential equation   for a general  time-dependent phase-space observable. 
 The GLE is derived from the Hamiltonian for an arbitrarily complex many-body system
%While we use  the GLE derived from classical Hamilton dynamics, a quantum version exists and can be used if the 
%observable exhibits quantum properties \cite{nakajima1958quantum}, which, however, is not the case for the systems considered here.
%
and   generalizes  the  Newtonian equation of motion by explicitly accounting for 
the coupling of the observable to its responsive environment in terms of non-Markovian  friction  and a time-dependent force.
Thus, the GLE correctly accounts for the loss of information when projecting the high-dimensional phase-space dynamics
onto the low-dimensional observable dynamics and constitutes an exact coarse-graining method.
Several methods to extract all GLE parameters from time-series data exist 
\cite{straub1987,daldrop_butane_2018, kowalik2019, grogan_data-driven_2020}.
 The GLE has been applied to  bond-length vibrations, dihedral rotations,   chemical reactions in solvents and protein folding
%\cite{satija2019generalized,ayaz2021non, brunig2022time, brunig2022pair,vroylandt2022likelihood,dalton2022protein,dalton2024role},
\cite{satija2019,ayaz2021,brunig2022a,brunig2022d,dalton2023,dalton2024}
 and also more complex  systems, such as the motion of living organisms and financial and meteorological data
 \cite{schmitt_analyzing_2006, mitterwallner2020, hassanibesheli2020reconstructing,klimek2024}. %, wand2023memory}. 
%Time-series are always discretized, 
%which poses severe problems for the applicability of the GLE, which is a time-continuous integro-differential equation \cite{niemann2008usage, tepper2024accurate}. 
%Thus, while the  GLE is universally applicable to arbitrary   systems and   in principle provides a suitable  framework for  time-series
%modeling, it was actually never used to forecast real-world complex data. This is due to four  complications:
%i) The multi-scale coupling between short-time stochastics and long-time transient or seasonal effects in the data, ii) the presence of non-Gaussian correlations in many 
%non-trivial systems, iii) time-discretization effects, and iv) the possible presence of non-equilibrium effects that render many tools of statistical mechanics useless. We solve all these problems by our combined filtering-projection approach.
But the problem of data  time-scale multiplicity is also encountered when using the GLE, mostly in the form 
of  numerical convergence problems when   extracting the GLE parameters from trajectories.

 We here introduce convolution filtering in conjunction with projection   as a method to deal with time-scale multiplicity problems  in 
time series data.  One central result is that a convolution-filtered trajectory still obeys Liouville dynamics and thus can be treated 
by exact phase-space projection techniques to  derive the GLE for filtered observables and to
 reveal  how the GLE parameters  are modified by filtering. 
 
 There are diverse applications for our filter-projection approach that are graphically illustrated in Fig. 1:
 i) For constructing  coarse-grained models,  a low pass filter removes irrelevant fast dynamics of an observable. 
 Our filter-projection approach  yields the coarse-grained
 effective parameters of the GLE (i.e. mass, friction and  potential) that exactly describe the long-time observable motion.
 ii) Conversely, every experimental measurement  entails some type of smoothening caused by the 
 finite time response of the measuring apparatus. Our filter projection approach can be used to invert the smoothening and calculate the 
 parameters of the original dynamic process if the filter characteristics are known.
 iii) For data that is plagued by slow transient or seasonal (i.e. periodic) effects, our filter-projection approach can be used 
 to split the data into the slow  and periodic components, which  can be treated by  simple deterministic models,
 and the fast components, which  include stochastic effects and for which the GLE approach is  ideally suited. 
Our filter-projection methodology  thus  is an alternative  to   non-equilibrium theoretical approaches towards
 the  dynamics of systems with slow transient dynamics.

\begin{figure*}	
	\centering
	\includegraphics[width=18cm]{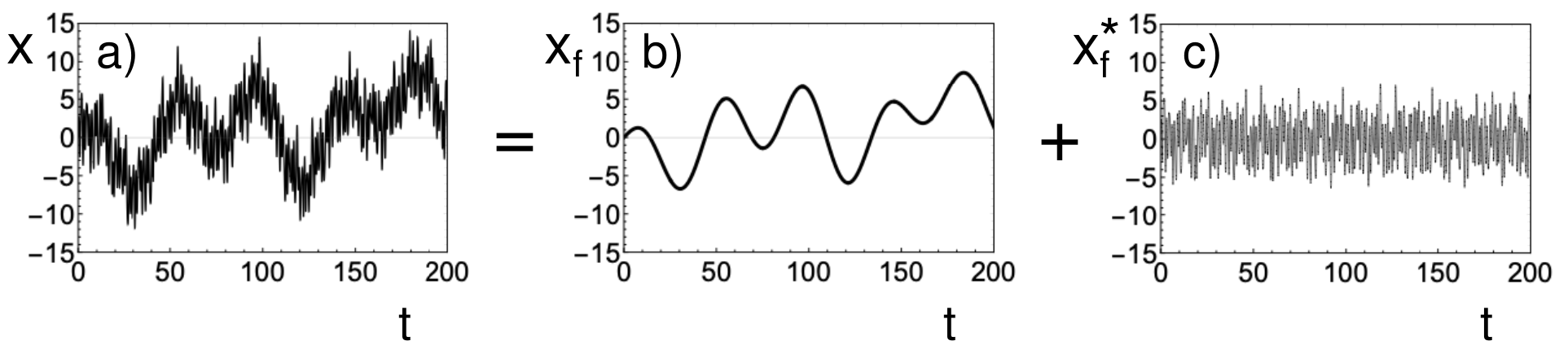}
\caption{ Graphical  illustration of the effect of filtering on a trajectory with multiple time scales. 
The trajectory $x(t)$ in a) is by convolution filtering decomposed into the low-pass filtered part
$x_f(t)$ in b) and the high-pass filtered complement $x_f^*(t)$ in c) such that  $x(t)=x_f(t) + x_f^*(t)$.
In coarse-graining applications one seeks the model that correctly describes the  stochastic and equilibrium  properties of the filtered component
in b). In data-reconstruction applications one would like to  infer the dynamics
of the unfiltered trajectory  in a) from the measured low-pass filtered trajectory in b). 
Finally, when dealing with systems that contain slow transient dynamics that cannot be fully resolved, the objective is to 
build models for the high-pass filtered component $x_f^*(t)$ in c). Our filter-projection approach applies  to all these scenarios. 
		}
	\label{fig1}
\end{figure*}

\section{The  generalized Langevin equation for filtered  observables  \label{GLE}}

\subsection{ From the Liouville equation to Heisenberg observables}

We consider a general  time-independent  Hamiltonian for  a system of $N$ interacting particles or atoms  in three-dimensional space
\begin{align}
\label{eq_Hamiltonian}
H (\omega) &= \sum_{n=1}^N \frac{\ve{p}_n^2}{2 m_n} + V ( \{ \ve r_N\}),
\end{align}
where a point in the 6$N$-dimensional  phase space  is denoted by 
$\omega = ( \{ \ve r_N\},  \{ \ve p_N\}  )$, which is a $6N$-dimensional  vector consisting
of the Cartesian particle positions  $\{ \ve r_N\}$ 
 and the  conjugate momenta  $ \{ \ve p_N\}$ and fully specifies the microstate of the system.
The Hamiltonian  splits into a kinetic and a potential part and  $m_n$ is the mass of particle  $n$. 
The potential $V ( \{ \ve r_N\})$ contains all interactions between the particles and 
includes possible external potentials. 

Using  the Liouville operator
\begin{align}
\label{eq_Liouville}
{\cal L}(\omega)&= \sum_{n=1}^N  \sum_{\alpha=x,y,z}  
\left(   \frac{ \partial H }{\partial p_n^\alpha }  \frac{ \partial  }{\partial r_n^\alpha }  -
 \frac{ \partial H }{\partial r_n^\alpha }  \frac{ \partial  }{\partial p_n^\alpha } 
 \right),
\end{align}
the  $6N$-dimensional  Hamilton equation of motion can be compactly written as
%\begin{align}
%\label{eq_time_evolution}
$\dot{\omega}(t) = {\cal L}(\omega)  \omega(t)$,
%\end{align}
where $\omega(t)$ is the  phase-space  location of the system at time $t$ 
and $\dot{\omega}(t) = {\rm d} \omega(t) /   {\rm d} t$  is the corresponding phase-space velocity.
From the Hamilton equation  and since the Liouville operator is time-independent, 
it follows that the phase space position  is propagated in time by the   operator exponential 
$e^{(t-t_0){\cal L}(\omega)}$, i.e., $e^{(t-t_0){\cal L}(\omega)}\omega(t_0) = \omega(t)$ \cite{zwanzig_nonequilibrium_2001}.
%\begin{subequations}
%\begin{align}
%\Phi(t): \Omega &\to \Omega\\
%\omega_0 &\mapsto \omega_t\\
%\Phi(t) &= e^{tL}.
%\end{align}
%\end{subequations}
Instead of following  microstate trajectories in phase space, which is the Lagrangian description of the system dynamics,
it is convenient to switch to the Eulerian description and
consider the time-dependent probability density distribution as a function of phase space, $\rho(\omega,t)$. 
The density  obeys the 
Liouville equation $\dot{\rho}(\omega,t) = - {\cal L} (\omega)  \rho(\omega,t)$,
which  has the formal solution $ \rho(t) = e^{-(t-t_0){\cal L}(\omega)}\rho(\omega,t_0)$
with some initial distribution $\rho(\omega,t_0)$, which need not be a stationary distribution \cite{zwanzig_nonequilibrium_2001}.
A system observable can be generally written as a Schrödinger-type (i.e. time-independent) phase-space function $B_{S}(\omega)$,
it  can for example represent  the position of one particle, its momentum, 
the center-of-mass position of a group of particles, or the reaction coordinate describing  a chemical reaction or the  folding of a protein.
Since the phase space can be defined as including the entire universe, so that no external time-dependent external perturbation needs 
to be taken into account  \cite{schmitt_analyzing_2006,Netz2024},  $B_{S}(\omega)$ can also represent more complex  observables such as the position or state of a living organism, 
a meteorological or an economical observable. 
 To simplify the notation,  we consider a scalar observable but note that 
our  formalism can be straightforwardly extended to multidimensional observables. 
Using the probability  density, the time-dependent expectation value (or mean) of the observable $B_{S}(\omega)$ can be written as
\begin{align}
\label{eq_mean}
b(t) &  \equiv \int {\rm d} \omega  \, B_{S}(\omega) \rho(\omega,t) \nonumber \\
&=  \int {\rm d} \omega  \, B_{S}(\omega) e^{-(t-t_0){\cal L}(\omega)}\rho(\omega,t_0).
\end{align}
Since the Liouville operator is anti-self adjoint, it follows that \cite{zwanzig_nonequilibrium_2001}
\begin{align}
\label{eq_mean2}
b(t) &=  \int {\rm d} \omega  \, \rho(\omega,t_0)  e^{(t-t_0){\cal L}(\omega)} B_{S}(\omega)   \nonumber \\
& = \int {\rm d} \omega  \, \rho(\omega,t_0)   B(\omega, t).
\end{align}
In the last step we have defined the Heisenberg observable as 
\begin{align}
\label{eq_Heisenberg}
B(\omega, t)  &\equiv  e^{(t-t_0){\cal L}(\omega)}  B_{S}(\omega),
\end{align}
which is the central object of the projection formalism and of GLEs.
Obviously, as follows from Eq.~\eqref{eq_Heisenberg}, 
it satisfies the equation of motion 
\begin{align}
\label{eq_Heisenberg2}
\dot{B}(\omega, t) & = {\cal L}(\omega) B(\omega, t)
\end{align}
with the initial condition    $B(\omega, t_0)=B_{S}(\omega)$.
To understand the meaning of a Heisenberg observable, we for the moment consider the 
density distribution $\rho(\omega,t_0)=\delta(\omega-\omega_0)$,
which describes a system that at time $t_0$ is in the microstate $\omega_0$. 
Inserting this into Eq.~\eqref{eq_mean2},  we obtain $b(t) = B(\omega_0, t)$.
In other words, $B(\omega_0, t)$ describes the dynamics  of an observable for a system  that at time $t=t_0$ was 
in the microstate $\omega_0$, i.e., it describes the temporal evolution of the conditional mean of the  observable  $B_{S}(\omega)$. 
It transpires that if we  derive an equation of motion for $B(\omega, t)$,
we have an equation for how this conditional mean changes in time.  
This is the central idea of projection and of GLEs.

\subsection{ Convolution-filtered Heisenberg observables} \label{sect_filterHeisenberg}

In all  what follows,  we will omit the phase-space argument of the Liouville operator and simply denote it as ${\cal L}$.
Introducing the general convolutional   filter function $f(t)$, we define the filtered mean of an observable as 
\begin{align}
\label{eq_meanfilter3}
b_f(t) &  \equiv  \int_{-\infty}^{\infty} {\rm d} s \, f(s)   b(t-s).
\end{align}
Filtering necessarily occurs whenever an experimental measurement is done, 
but it can also be used purposely to
decompose data into separate contributions, to
 remove unwanted components of time series data 
or to enhance certain interesting features. Using Eq.~\eqref{eq_mean2},
the time filtering of the observable mean can be written in terms of a phase-space average as
\begin{align}
\label{eq_meanfilter4}
b_f(t) 
&=  \int_{-\infty}^{\infty} {\rm d} s \, f(s)  \int {\rm d} \omega  \, \rho(\omega,t_0)   B(\omega, t-s) \nonumber \\
&=  \int {\rm d} \omega  \,  \rho(\omega,t_0)   \int_{-\infty}^{\infty} {\rm d} s f(s)  B(\omega, t-s) \nonumber \\
&=  \int {\rm d} \omega  \,  \rho(\omega,t_0)   B_f(\omega, t ).
\end{align}
In the last line we defined the filtered Heisenberg observable 
\begin{align}
\label{eq_meanfilter5}
 B_f(\omega, t) & \equiv   \int_{-\infty}^{\infty} {\rm d} s \, f(s)  B(\omega, t-s)   \\
&=  \int_{-\infty}^{\infty} {\rm d} s \, f(s)  e^{(t-s-t_0) {\cal L}} B_{S}(\omega),
\end{align}
which by construction has the same properties as the unfiltered Heisenberg observable in 
Eqs.~\eqref{eq_Heisenberg} and \eqref{eq_Heisenberg2}: it is time propagated according to 
\begin{align}
\label{eq_Heisenfilter}
e^{t'  {\cal L}} B_f(\omega,t) &= e^{t'{\cal L}}  \int_{-\infty}^{\infty} {\rm d} s \, f(s)  
e^{(t-s-t_0) {\cal L}} B_{S}(\omega)\nonumber \\
&= \int_{-\infty}^{\infty} {\rm d} s \, f(s)  e^{(t'+t-s-t_0) {\cal L}} B_{S}(\omega) \nonumber \\
&= B_f(\omega,t'+t)
\end{align}
and it satisfies the equation of motion 
\begin{align}
\label{eq_Heisenfilter2}
\dot{B}_f(\omega, t) & = {\cal L} B_f(\omega, t).
\end{align}
Thus, although $B_f(\omega,t)$ involves a weighted integral over time and thus is not an ordinary 
 observable, i.e., an  instantaneously defined phase-space function, 
its dynamics is identical to that of an ordinary Heisenberg observable. 
This simple but profound property, which is one of the main results of this paper,
 will allow us to derive a GLE for the filtered Heisenberg observable $ B_f(\omega, t)$ in exactly
the same way as it is commonly done  for the ordinary (unfiltered)  Heisenberg observable $ B(\omega, t)$.

\subsection{Convolution filter examples}

In principle any function $f(t)$ can be employed as a temporal filter, but there are a few particularly useful
 filter functions. 
In Fourier space Eq.~\eqref{eq_meanfilter5} reads 
\begin{align}
\label{eq_filter1}
 \tilde{B}_f(\omega, \nu) & =   \tilde{f}(\nu)   \tilde{B}(\omega, \nu),
\end{align}
where we define Fourier transforms as $  \tilde{f}(\nu)   = \int {\rm d} t e^{-\imath \nu t} f(t)$. 
The normalized Gaussian filter with a temporal width $\lambda$,
\begin{align}
\label{eq_filter2}
 f_{\rm LP} (t)  & =   e^{-t^2/(2 \lambda^2)} (2 \pi \lambda^2)^{-1/2} ,
\end{align}
is a low-pass filter that is commonly used for  smoothing experimental and simulation data.
Its Fourier transform is given by a Gaussian as well,
\begin{align}
\label{eq_filter3}
 \tilde{f}_{\rm LP} (\nu)  & =   e^{- \lambda^2 \nu^2 /2}.
\end{align}
The  Gaussian high-pass filter is the complement of the Gaussian low-pass filter and is given by
\begin{align}
\label{eq_filter4}
 f_{\rm HP} (t)  & = \delta(t) -  f_{\rm LP} (t) 
\end{align}
with the  Fourier transform 
\begin{align}
\label{eq_filter5}
 \tilde{f}_{\rm HP} (\nu)  & = 1 -    e^{- \lambda^2 \nu^2 /2}.
\end{align}
It can be used to eliminate the slow or transient dynamics of a trajectory. 
The  normalized Gaussian band-pass filter
\begin{align}
\label{eq_filter6}
 f_{\rm BP} (t,\nu_0)  & = \frac{ \cos(\nu_0 t)   e^{-t^2/(2 \lambda^2)} }
 { (2 \pi \lambda^2)^{1/2}  e^{- \lambda^2 \nu_0^2/2}}
\end{align}
with its Fourier transform 
\begin{align}
\label{eq_filter5b}
 \tilde{f}_{\rm BP} (\nu, \nu_0)  & =  \frac{ e^{- \lambda^2 ( \nu- \nu_0)^2/2 } + e^{- \lambda^2 ( \nu+ \nu_0)^2/2 } }
 {2 e^{- \lambda^2 \nu_0^2/2}}
\end{align}
can be used to perform  convolutional Fourier analysis. To illustrate its properties, we consider  as an example
  a cosine  function with a finite phase $\phi$, 
  \begin{align}
B(\omega,t)= B_0(\omega)  \cos(\nu'_0 t+ \phi),
\end{align}
where $B_0(\omega)$ is an arbitrary phase-space function. $B(\omega,t)$
has the Fourier transform
  \begin{align}
\tilde{B}(\omega,\nu)= \pi B_0(\omega)  
\left( e^{\imath \phi} \delta(\nu - \nu'_0) + e^{-\imath \phi} \delta(\nu + \nu'_0)\right).
\end{align}
According to Eqs.~\eqref{eq_filter1}  and \eqref{eq_filter6} and after a few intermediate steps, 
the filtered function can be written in Fourier space as
  \begin{align}
\tilde{B}_f(\omega,\nu,\nu_0)=
\tilde{B}(\omega,\nu)
\frac{e^{- \lambda^2 (\nu_0-\nu'_0)^2/2 } +e^{- \lambda^2 (\nu_0+\nu'_0)^2/2 }}
{ 2 e^{- \lambda^2 \nu_0^2/2}}
\end{align}
and becomes sharply peaked for a filter frequency $\nu_0$ 
around the frequency $\nu'_0$ of the input function 
in the limit $\lambda\nu'_0 \gg 1$.
This simple example shows that the Gaussian convolutional band-pass filter projects
 onto oscillatory contributions in time-series data and thus can be used analogously 
 to ordinary Fourier transformation (which, obviously, does not correspond to  a convolution). 
The  Gaussian band-stop filter is the complement of the Gaussian band-pass filter and given by
\begin{align}
\label{eq_filter6b}
 f_{\rm BS} (t, \nu_0)  & = \delta(t) -  f_{\rm BP} (t) 
\end{align}
with the  Fourier transform 
\begin{align}
\label{eq_filter7}
 \tilde{f}_{\rm BS} (\nu, \nu_0)  & = 1 -  
 \frac{ e^{- \lambda^2 ( \nu- \nu_0)^2/2 } + e^{- \lambda^2 ( \nu+ \nu_0)^2/2 } }
 {2 e^{- \lambda^2 \nu_0^2/2}} .
 \end{align}
It can be used to eliminate  oscillatory  dynamics with oscillation 
frequency $\nu_0$  in time-series data. Of course, one can apply an arbitrary combination
of filter functions onto time-series data and the resulting function behaves as a regular 
phase-space observable, as explained in Sect. \ref{sect_filterHeisenberg}.

\subsection{ Derivation of the GLE for convolution-filtered Heisenberg observables}

Using the definition Eq.~\eqref{eq_meanfilter5}, we  decompose the unfiltered 
Heisenberg observable $ B(\omega, t)$  into the filtered Heisenberg observable
 $B_f(\omega, t)$ and its complement $B^*_f(\omega, t)$ according to 
\begin{align}
\label{eq_decomp}
 B(\omega, t) &= B_f(\omega, t)+  B^*_f(\omega, t).
\end{align}
Note that this decomposition is exact and can be viewed as the definition of  the complement. 
By adding many convolution-filtered Heisenberg observables according to 
$B_f(\omega, t) = \sum_i B^i_f(\omega, t)$, where the $i$ numbers different convolution
filters, we can apply multiple  low-pass, high-pass, band-pass and band-stop filters
 on the data. 
 According to Eq.~\eqref{eq_decomp}, 
 the acceleration of the unfiltered  Heisenberg observable can be decomposed as 
\begin{align}
\label{eq_decomp2}
\ddot B(\omega, t) &= \ddot B_f(\omega, t)+  \ddot B^*_f(\omega, t).
\end{align}
As a matter of fact, one can analyze the two parts on the right side
of  Eq.~\eqref{eq_decomp2}  separately and  using different types of theoretical description. 
For the  example of a high-pass filter, one could use a stochastic description for the high-pass filtered part $B_f(\omega, t)$,
which contains   the fast fluctuations, 
while the low-pass complement $B^*_f(\omega, t)$, which contains  the slow components,  could be described by a deterministic model.
In the following we will apply the  projection formalism  on 
$\ddot B_f(\omega, t)$ but stress that projection  can be equally well  applied  on the complementary part
$\ddot B_f^*(\omega, t)$.

Here we follow standard procedures \cite{zwanzig_nonequilibrium_2001}. 
We introduce a projection operator ${\cal P}$ that acts on phase space functions
 and its complementary operator ${\cal Q}$ via the relation $1= {\cal Q} + {\cal P}$.
 Inserting this unit operator into the time propagator relation Eq.~\eqref{eq_Heisenfilter} 
and using Eq.~\eqref{eq_Heisenfilter2}, we obtain 
\begin{align}
\label{eq_GLE1}
 & \ddot B_f(\omega, t)= e^{(t-t_P)  {\cal L}} ( {\cal P}+ {\cal Q}) {\cal L}^2  B_f(\omega,t_P)  \nonumber \\
&= e^{(t-t_P)  {\cal L}}  {\cal P} {\cal L}^2  B_f(\omega,t_P) + e^{(t-t_P)  {\cal L}}  {\cal Q} {\cal L}^2  B_f(\omega,t_P), 
\end{align}
where $t_P$ defines the time at which the projection is performed, which in principle can 
differ from the time $t_0$ at which the time propagation of the  Heisenberg 
variable in Eq.~\eqref{eq_Heisenfilter} starts.
% (the relevance of this will become clear later).
The projection operators ${\cal Q}$ and  ${\cal P}$ in general depend on the projection time $t_P$ 
(which is not explicitly written out) but not on the time $t$.
 This allows us to use the  standard Dyson operator 
decomposition \cite{dyson_radiation_1949, feynman_operator_1951, evans_statistical_2008} for
 the propagator $e^{(t-t_P)  {\cal L}}$
\begin{align}
\label{eq_GLE2}
e^{(t-t_P)  {\cal L}}= e^{(t-t_P)Q{\cal L}} + \int_0^{t-t_P} \mathrm{d}s\,e^{(t-t_P-s){\cal L}} {\cal P} {\cal L} e^{sQ {\cal L}}.
\end{align}
Inserting this decomposition into the second term on the right hand side  in Eq.~\eqref{eq_GLE1},
we obtain the GLE in general form 
\begin{align}
\label{eq_GLE3}
  \ddot B_f(\omega, t) & =  e^{(t-t_P)   {\cal L}}  {\cal P}  {\cal L}^2  B_f(\omega,t_P) + F(\omega,t) \nonumber \\
& + \int_0^{t-t_P} \rm{d}s\, e^{(t-t_P-s) {\cal L}} {\cal P}  {\cal  {\cal L}}  F(\omega,s+t_P),
\end{align}
where the complementary  force is given by
\begin{align}
\label{eq_F_operator}
F(\omega,t)&  \equiv  e^{(t-t_P) {\cal Q} {\cal L}} {\cal Q}{\cal L}^2  B_f(\omega, t_P).
\end{align}
The first term on the right-hand side in Eq.~\eqref{eq_GLE3}
will turn out to represent the  conservative force from a potential,
the third term represents non-Markovian friction effects  and the force $F(\omega,t)$ 
represents all effects that are not included in the other two terms.   
 $F(\omega,t)$  is a function of  phase space and evolves in the complementary space,
 i.e. it satisfies $ {\cal P} F(\omega,t)=0$ (as will be shown  further below). 

 Clearly, the explicit form of Eq.~\eqref{eq_GLE3} depends on the specific form of the projection operator 
 $\cal P$.  Here we choose the Mori projection, because it is  straightforward to implement
 and for the derivation of the filtering effects  based on two-point correlations functions is exact,  as we will discuss below.
 We note that our  filter-projection approach can also be used in conjunction with  hybrid  projection operators
 for which the resulting GLE  explicitly  contains a non-linear potential of mean force  \cite{Ayaz2022,vroylandt_epl_2022,Ayaz2022b}.
The Mori projection using $B_f(\omega, t_P)$ as a projection function and
applied on a general Heisenberg  observable $A(\omega,t)$ is given by \cite{mori_transport_1965,vroylandt_epl_2022}
\begin{align}
\label{eq_mori_projection}
 &{\cal P}  A(\omega,t) = \langle   A(\omega,t)  \rangle
  +  \frac{\langle   A(\omega,t)  {\cal L} B_f (\omega, t_P)  \rangle}  {\langle ( {\cal L}B_f(\omega, t_P) )^2 \rangle} 
  {\cal L}B_f (\omega, t_P)  \nonumber \\
 &  +\frac{\langle   A(\omega,t)  (B_f (\omega, t_P) - \langle B_f \rangle )\rangle}  {\langle (B_f (\omega, t_P) -  \langle B_f \rangle  )^2 \rangle} 
 (B_f (\omega, t_P) - \langle B_f \rangle ).
  \end{align}
We define the expectation value of an arbitrary phase-space function $X(\omega) $ 
  with respect to a time-independent projection  distribution 
$ \rho_{\rm P}(\omega)$ as
\begin{align} \label{projectdist}
\langle   X(\omega)  \rangle = 
\int {\rm d}\omega   X (\omega) \rho_{\rm P}(\omega),
  \end{align}
  which we here take to be the normalized equilibrium canonical distribution
    $\rho_{\rm P}(\omega)= e^{-H(\omega)/(k_BT)}/ Z$, where  $Z$ is the partition function. 
  For  this stationary projection  distribution the average
$\langle B_f \rangle  = \langle B_f(\omega,t_P) \rangle$ is independent of time.
The Mori  projection in Eq.~\eqref{eq_mori_projection} 
projects onto a constant,  the filtered observable $B_f (\omega, t_P)$ and its time derivative 
$ {\cal L}B_f (\omega, t_P)$ at time $t_P$, the projection time. 
Thus, the projection in Eq.~\eqref{eq_mori_projection} maps any observable $A(\omega,t)$
 onto the subspace of all functions linear in the observables 
 1, $B_f (\omega, t_P)$ and $ {\cal L}B_f (\omega, t_P)$, meaning that 
 ${\cal P}1=1$, ${\cal P}B_f (\omega, t_P)= B_f (\omega, t_P)$ and 
 ${\cal P}{\cal L}B_f (\omega, t_P)= {\cal L}B_f (\omega, t_P)$. From this follows immediately 
 that ${\cal Q}1= {\cal Q}B_f (\omega, t_P)={\cal Q}{\cal L}B_f (\omega, t_P)= 0$, which is
 a  property that will become important later on. 
 
The Mori projection is  linear,
 i.e., for two arbitrary observables $A(\omega,t)$ and $C(\omega,t')$ it satisfies
 ${\cal P} (c_1 A(\omega,t) + c_2 C(\omega,t'))=c_1 {\cal P} A(\omega,t)  + c_2 {\cal P} C(\omega,t')$,
 it is idempotent, i.e.,  ${\cal P}^2= {\cal P}$, and it is self-adjoint, i.e. it satisfies 
  the relation
\begin{align}
\langle A(\omega,t) {\cal P} C(\omega,t')  \rangle = \langle C(\omega,t') {\cal P} A(\omega,t)   \rangle.
\label{eq_projection_orthogonal}
\end{align} 
From these properties it follows that the complementary  projection
 operator ${\cal Q}=1-{\cal P}$ is also linear,  idempotent and self-adjoint.
Thus, ${\cal P}$ and ${\cal Q}$ are orthogonal to each other, i.e. 
${\cal P}{\cal Q}= 0 = {\cal Q}{\cal P}$ \cite{Netz2024}. 

Using all these properties, the GLE  Eq.~\eqref{eq_GLE3} takes the form \cite{mori_transport_1965, zwanzig_nonequilibrium_2001}
\begin{align} \label{eq_mori_GLE}
  \ddot B_f(\omega, t) & = -  K_f   (B_f(\omega,t) - \langle B_f \rangle )  + F(\omega,t) \nonumber \\
& - \int_0^{t-t_P} {\rm d}s\, \Gamma_f(s) \dot B_f(\omega, t-s),
\end{align}
details of the derivation are shown in Appendix \ref{sec_App_GLE}.
The parameter $K_f$, which corresponds to the potential stiffness  divided by the effective mass, 
 is given by 
\begin{align} \label{GLEK}
K_f &= \frac{\langle ( {\cal L}B_f(\omega, t_P) )^2 \rangle}
{\langle (B_f (\omega, t_P) - \langle B_f \rangle )^2 \rangle} 
\end{align}
and the memory friction kernel is given by 
\begin{align} 
\Gamma_f(s) = 
  \frac{\langle F(\omega,t_P) F(\omega, s+t_P ) \rangle}{\langle ( {\cal L}B_f(\omega, t_P) )^2 \rangle}=
\frac{\langle F(\omega,0) F(\omega, s) \rangle}{\langle ( {\cal L}B_f(\omega, t_P) )^2 \rangle}.
 \label{eq_mori_memory}
\end{align}
 Eq.~\eqref{eq_mori_GLE} is an exact decomposition of the Liouville equation into three terms, the first term is a force 
 due to a quadratic potential, the third term accounts for linear non-Markovian  friction and depends on the memory kernel $\Gamma_f(s)$,
  which is related via Eq.~\eqref{eq_mori_memory} to the second moment of the  force $F(\omega,t)$, 
defined in Eq.~\eqref{eq_F_operator}.
In particular,  eq.~\eqref{eq_mori_GLE} is exact and time-reversible, as are the underlying Hamilton 
and Liouville equations.
The explicit  form of the memory function can  be computed for simple model systems,
as will be demonstrated later on. 
Note that the  force $F(\omega,t)$  can
  be extracted from simulation or experimental data by different techniques
 \cite{carof_two_2014, lesnicki_molecular_2016,Ayaz2022}. 
 Due to the specific form of the Mori projection in Eq.~\eqref{eq_mori_projection},
 several expectation values involving  the   force vanish, namely 
  $\langle F(\omega,t)\rangle = \langle   F(\omega,t)B_f(\omega, t_P) \rangle =
  \langle   F(\omega,t){\cal L}B_f(\omega, t_P) \rangle =0$, which 
  are important properties for extracting GLE parameters from time-series data.

 For practical applications, one typically models the  force $F(\omega,t)$ as a stochastic process with 
 zero mean and a second moment given by Eq.~\eqref{eq_mori_memory},
  higher-order moments of $F(\omega,t)$ 
  are typically neglected and the distribution of $F(\omega,t)$ is  assumed to be Gaussian, which is 
  valid for  the calculation of two-point correlation functions. 
 For non-linear  systems and if one is interested in higher-order correlations, however, this assumption can not hold, 
 since $F(\omega,t)$ is the only term in the GLE that accounts for   non-linearities. 
 Thus, imposing $F(\omega,t)$ to be a Gaussian variable  can become a bad approximation for non-linear systems,
 which reflects a fundamental short-coming of the Mori projection scheme in conjunction with 
 replacing $F(\omega,t)$ by a random Gaussian process.
 Alternative methods to derive GLEs with non-linear potential and friction terms have been recently proposed
  \cite{Ayaz2022,vroylandt_epl_2022,vroylandt_likelihood_2022,Ayaz2022b}.
 We stay here with the Mori scheme because it simplifies analytical calculations and 
 is exact for our calculations of two-point correlations. We note that
  high-pass filtering  produces data that is Gaussian to a very good approximation. 
 
 \section{Dynamic properties of  filtered  trajectories}
 
 \subsection{Parameters of the  GLE for filtered observables}
 
 We now interpret the  GLE in Eq.~\eqref{eq_mori_GLE} for the 
 phase-space dependent filtered Heisenberg observable $B_f(\omega, t)$ as  a stochastic 
 equation. For a more stream-lined notation, 
 we skip the  phase-space dependence and
 define the filtered observable  by subtracting its mean as 
 \begin{align} 
x_f(t) = B_f(\omega, t) -  \langle B_f \rangle .
\end{align}
We also shift the projection time into the far past, $t_P \rightarrow - \infty$,  define
the memory kernel to be causal or single-sided, i.e. $\Gamma_f(t)=0$ for $t<0$, 
and define a stochastic random  force as  $F_R(t) = F(\omega,t)$.
%(details of this non-trivial transformation are given in Appendix \ref{sec_App_GLEx}).
The GLE %becomes time-homogeneous and 
reads 
 \begin{align} \label{eq_mori_GLE2}
  \ddot x_f(t) & = -  K_f   x_f(t)  + F_R(t)
 - \int_{-\infty}^{\infty} {\rm d}s\, \Gamma_f(s) \dot x_f(t-s),
\end{align}
where the potential  parameter is, following Eq.~\eqref{GLEK},  given by 
\begin{align}  \label{K} 
K_f &= \frac{\langle \dot x_f^2 \rangle}
{\langle x_f^2 \rangle} 
\end{align}
and the memory friction kernel is, following Eq.~\eqref{eq_mori_memory},  given by 
\begin{align} 
\Gamma_f(t) =  \theta(t)
 \frac{\langle F_R(0) F_R( t) \rangle}{\langle  \dot x_f^2 \rangle}
 \label{eq_mori_memory3}
\end{align}
where $ \theta(t)$ defines the Heavyside function
and the mean-squared velocity $ \langle \dot x^2 \rangle$
would, for a standard equilibrium observable and  according to the equipartition theorem,
 correspond   to the thermal energy  divided by the effective mass.
The  force $F_R(t)$  is characterized  by its second
moment according to Eq.~\eqref{eq_mori_memory3}, while its first moment  vanishes, $\langle F_R(t) \rangle=0$,
as explained before. This, however, does not mean that the random force is necessarily Gaussian.
Note that from now on,  all expectation values are defined as averages with respect to the random force  distribution. This 
reinterpretation of the GLE in Eq.~\eqref{eq_mori_GLE} as a stochastic integro-differential 
equation   simplifies the notation significantly and is exact as long as care is taken to relegate
all phase-space dependencies into the random force $F_R(t)$. 

By Fourier transformation, Eq.~\eqref{eq_mori_GLE2} can be solved in terms of the linear-response relation as
\begin{align}  \label{eq_response}
\tilde x_f(\nu) = \tilde \chi_f(\nu) \tilde F_R(\nu),
\end{align}
where the response function is given as 
\begin{align}  \label{eq_response2}
\tilde \chi_f(\nu)  = \left[ K_f - \nu^2 + \imath \nu \tilde \Gamma_f(\nu) \right]^{-1}. 
\end{align}
We define the filtered  two-point positional auto-correlation function as 
\begin{align} 
C_{xx}^f(t) = \langle x_f(t') x_f(t+t') \rangle,
\end{align}
the Fourier transform of which reads 
\begin{align}  \label{eq_Cxx1}
&\tilde C_{xx}^f(\nu) = \int \frac{{\rm d}t'}{2\pi}   e^{\imath t'(\nu+\nu')} 
 \langle \tilde x_f(\nu) \tilde x_f(\nu')   \rangle \nonumber \\
& = \int \frac{{\rm d}t'}{2\pi}   e^{\imath t'(\nu+\nu')}  \tilde \chi_f(\nu) \tilde \chi_f(\nu')
   \langle \tilde F_R(\nu) \tilde F_R(\nu')   \rangle 
\end{align}
where in the last step we used Eq.~\eqref{eq_response}.
The random-force autocorrelation in Fourier space follows from Eq.~\eqref{eq_mori_memory3}
as 
\begin{align} 
 \langle \tilde F_R(\nu) \tilde F_R( \nu') \rangle =
 2\pi \delta(  \nu+\nu') \langle  \dot x_f^2 \rangle
\left[ \tilde \Gamma_f(\nu) + \tilde \Gamma_f(\nu')\right].
   \label{eq_mori_memory4}
\end{align}
Combining Eqs.~\eqref{eq_Cxx1} and \eqref{eq_mori_memory4} we obtain
\begin{align}  \label{eq_Cxx2}
&\tilde C_{xx}^f(\nu)= \langle  \dot x_f^2 \rangle  \tilde \chi_f(\nu) \tilde \chi_f(-\nu) 
\left[ \tilde \Gamma_f(\nu) + \tilde \Gamma_f(-\nu)\right]  \nonumber \\
&= \frac{ \langle  \dot x_f^2 \rangle}{\imath \nu} 
\left[ \tilde \chi_f(-\nu) - \tilde \chi_f(\nu)  \right],
\end{align}
where in the last step we used Eq.~\eqref{eq_response2}.
Eq.~\eqref{eq_Cxx2} forms the starting point 
for a few further derivations and is exactly produced by the Mori GLE. 
First, we multiply  Eq.~\eqref{eq_Cxx2} by $\imath \nu$ to obtain
\begin{align}  \label{eq_Cxx3}
 \tilde C_{x\dot x}^f(\nu) \equiv \imath \nu \tilde C_{xx}^f(\nu)=  \langle  \dot x_f^2 \rangle 
\left[ \tilde \chi_f(-\nu) -\tilde \chi_f(\nu)  \right],
\end{align}
where $ \tilde C_{x\dot x}^f(\nu)$ is the Fourier transform of the derivative of the 
positional autocorrelation function $C_{x\dot x}^f(t) \equiv  {\rm d}C_{xx}^f(t)/{\rm d}t= \langle x_f(0) \dot x_f(t)\rangle$.
The right-hand side of Eq.~\eqref{eq_Cxx3} splits into a causal part, $\tilde \chi_f(\nu)$, and an
anticausal part, $\tilde \chi_f(-\nu)$, therefore we can also split the left-hand side into causal and anticausal
parts and finally obtain 
\begin{align}  \label{eq_Cxx4}
  \tilde C_{x\dot x}^{+f}(\nu)= -  \langle  \dot x_f^2 \rangle
\tilde \chi_f(\nu),
\end{align}
where $ \tilde C_{x\dot x}^{+f}(\nu)$ is the Fourier transform of the single-sided correlation function,
$C_{x\dot x}^{+f}(t)= \theta(t) C_{x\dot x}^f(t)$. % and $ \theta(t)$ is the Heaviside function. 
Eq.~\eqref{eq_Cxx4} is the Fourier transform of the fluctuation dissipation theorem \cite{zwanzig_nonequilibrium_2001} and one can use 
it to extract all parameters of the filtered GLE from  filtered time series data. To see this we combine 
Eqs.~\eqref{eq_Cxx4} and \eqref{eq_response2} to obtain
\begin{align}  \label{eq_Cxx5}
 - \frac{ \langle  \dot x_f^2 \rangle }{ \tilde C_{x\dot x}^{+f}(\nu)}= 
K_f - \nu^2 - \nu \tilde \Gamma_f''(\nu)  + \imath \nu \tilde \Gamma_f'(\nu),
\end{align}
 where we split the memory function into its real and imaginary parts according to 
 $\tilde \Gamma_f(\nu)=  \tilde \Gamma_f'(\nu)  + \imath \tilde \Gamma_f''(\nu)$.
 It transpires that the real and imaginary parts of the memory function are determined by
 \begin{subequations}  \label{eq_Cxx6}
\begin{align} 
\Gamma_f'(\nu)=  - \left( \frac{ \langle  \dot x_f^2 \rangle }{\nu \tilde C_{x\dot x}^{+f}(\nu)}\right)'',
\\
\Gamma_f''(\nu)= \frac{K_f}{\nu} -\nu+
 \left( \frac{ \langle  \dot x_f^2 \rangle }{\nu  \tilde C_{x\dot x}^{+f}(\nu)}\right)'.
\end{align}
\end{subequations}
The potential parameter $K_f$ follows from  Eq.~\eqref{K} or, alternatively, from Eq.~\eqref{eq_Cxx5} in the zero-frequency limit as 
\begin{align}  \label{eq_Cxx7}
K_f= - \frac{ \langle  \dot x_f^2 \rangle }{ \tilde C_{x\dot x}^{+f}(0)} = 
\frac{ \langle  \dot x_f^2 \rangle }{ \langle  x_f^2 \rangle },
\end{align}
where we used that  the Fourier transform of the memory function is finite in the zero frequency limit.
 %and inserted eq.~\eqref{K}. 
 In numerical applications,  Eq.~\eqref{eq_Cxx7} can serve  as a  check 
on the accuracy  of the numerical determination of  $\tilde C_{x\dot x}^{+f}(\nu)$.
Eqs.~\eqref{eq_Cxx6} and \eqref{eq_Cxx7} can be used to extract all parameters of the 
filtered GLE in Eq.~\eqref{eq_mori_GLE2} from a filtered  trajectory, since $\tilde C_{x\dot x}^{+f}(\nu)$ and 
 the expectation values $ \langle  x_f^2 \rangle$ 
and $\langle  \dot x_f^2 \rangle$ can be directly evaluated from time-series  data.

Let us derive a second useful result from Eq.~\eqref{eq_Cxx2}. Since 
$\chi_f(t)$ is a real function, it follows that  $\tilde \chi_f'(\nu)= \tilde \chi_f'(-\nu)$
and  $\tilde \chi_f''(\nu)=- \tilde \chi_f''(-\nu)$. Using this in Eq.~\eqref{eq_Cxx2},
we obtain 
\begin{align}  \label{eq_Cxx8}
&\tilde C_{xx}^f(\nu)= 
- \frac{2  \langle  \dot x_f^2 \rangle \tilde \chi_f''(\nu) }{ \nu},
\end{align}
which is an alternative formulation of the fluctuation-dissipation theorem in Fourier space \cite{zwanzig_nonequilibrium_2001}.
In the zero-frequency limit we obtain from  Eq.~\eqref{eq_Cxx8} 
\begin{align}  \label{eq_Cxx9}
&\tilde C_{xx}^f(0)= 
 \frac{2  \langle  \dot x_f^2 \rangle \tilde \Gamma_f(0) }{ K_f^2} =
 \frac{2  \langle x_f^2 \rangle \tilde \Gamma_f(0) }{ K_f},
\end{align}
where  we used Eqs.~\eqref{eq_response2} and \eqref{K}
and $\tilde \Gamma_f(0)$, the integral over the memory function, is the friction coefficient.
For unconfined systems with  $K_f=0$, $\tilde \Gamma_f(0)$ can be obtained from  Eq.~\eqref{eq_Cxx8} using the 
velocity autocorrelation function $C_{\dot x \dot x}^f(t) 
\equiv  \langle \dot x_f(0) \dot x_f(t)\rangle = -  {\rm d^2}C_{xx}^f(t)/{\rm d}t^2$ as 
\begin{align}  \label{eq_Cxx10}
&\tilde C_{\dot x \dot x}^f(0)= 
 \frac{2  \langle  \dot x_f^2 \rangle }{ \tilde \Gamma_f(0) },
\end{align}
which corresponds to  the standard  relation between the integral over the velocity autocorrelation function, $\tilde C_{\dot x \dot x}^f(0)$,
 and the diffusion constant  \cite{zwanzig_nonequilibrium_2001}.
Relations Eqs.~\eqref{eq_Cxx9} and \eqref{eq_Cxx10}  can be used to obtain the friction coefficient  $\tilde \Gamma_f(0)$  from filtered trajectory data. 

\subsection{Mapping between   filtered and  unfiltered GLE parameters} 

We now discuss the mapping between the filtered observable $x_f(t)$, described
by the GLE in Eq.~\eqref{eq_mori_GLE2}, and the unfiltered observable, denoted 
as $x(t)$ and which we describe by the same GLE but with different (i.e. unfiltered) 
parameters $K$, $\Gamma(t)$ 
and an unfiltered  random force that  is characterized by its second moment
according to Eq.~\eqref{eq_mori_memory3} 
where   $\langle \dot x^2_f \rangle$ is replaced by $\langle \dot x^2 \rangle$.
Similarly  to Eq.~\eqref{eq_filter1}, the relation between $x_f(t)$ and $x(t)$ is given by
\begin{align}
\label{eq_map1}
 \tilde{x}_f(\nu)  =   \tilde{f}(\nu)   \tilde{x}(\nu).
\end{align}
The question  is, given the parameters of the  GLE for the unfiltered observable, what are
the parameters of the GLE for the filtered observable, or vice versa. As turns out, this is 
a  rather  non-trivial question. 

First, using Eq.~\eqref{eq_map1},
the positional autocorrelation function for the filtered observable  $x_f(t)$ in Eq.~\eqref{eq_Cxx1}
can be written as 
\begin{align}  \label{eq_map2}
\tilde C_{xx}^f(\nu)& = \int \frac{{\rm d}t'}{2\pi}   e^{\imath t'(\nu+\nu')} 
\tilde{f}(\nu) \tilde{f}(\nu')  \langle \tilde x(\nu) \tilde x(\nu')   \rangle \nonumber \\
& = \tilde{f}(\nu) \tilde{f}(-\nu)\tilde C_{xx}(\nu),
\end{align}
where we followed the  same steps leading to Eq.~\eqref{eq_Cxx2}. This relation
 connects the autocorrelations for the filtered observable 
$\tilde C_{xx}^f(\nu)$ and  the unfiltered observable 
$\tilde C_{xx}(\nu)$ and  furnishes the central relation 
for understanding the effect of filtering on the system dynamics.

Combining Eqs.~\eqref{eq_Cxx9} and \eqref{eq_map2}
leads to 
\begin{align}  \label{eq_map3}
 \frac{ \langle \dot x_f^2 \rangle \tilde \Gamma_f(0) }{ K^2_f} = 
 \frac{\tilde{f}^2(0) \langle \dot x^2 \rangle \tilde \Gamma(0) }{ K^2},
\end{align}
which is a   relation between the parameters of the GLEs describing the 
filtered and unfiltered observables. 
In the following sections we will   disentangle the filtering effect on the GLE  parameters.

\subsubsection{Extracting filtered parameters from  unfiltered trajectories}

We now combine Eqs.~\eqref{eq_Cxx3} and \eqref{eq_map2} to obtain
\begin{align}  \label{eq_map4}
 \tilde \chi_f(-\nu) -\tilde \chi_f(\nu) &  = 
\frac{ \tilde{f}(\nu) \tilde{f}(-\nu)  \imath \nu \tilde C_{xx}(\nu)}
{ \langle  \dot x_f^2 \rangle } \nonumber \\
&\equiv  \frac{  \tilde C_{u\rightarrow f}(\nu) }
{ \langle  \dot x_f^2 \rangle }.
\end{align}
By defining the the single-sided  time-domain function 
$C^+_{u\rightarrow f}(t) = \theta(t) C_{u\rightarrow f}(t)$,
 where $C_{u\rightarrow f}(t)$ is the back Fourier transform of 
$\tilde C_{u\rightarrow f}(\nu) \equiv \tilde{f}(\nu) \tilde{f}(-\nu)  \imath \nu \tilde C_{xx}(\nu)$,
we obtain by splitting Eq.~\eqref{eq_map4} into its causal and anticausal parts
\begin{align}  \label{eq_map5}
 - \frac{ \langle  \dot x_f^2 \rangle }{ \tilde C_{u\rightarrow f}^{+}(\nu)}= 
 \tilde \chi_f^{-1} (\nu) = 
K_f - \nu^2 - \nu \tilde \Gamma_f''(\nu)  + \imath \nu \tilde \Gamma_f'(\nu).
\end{align}
Following the same strategy leading to Eq.~\eqref{eq_Cxx6} we obtain 
 \begin{subequations}  \label{eq_map6}
\begin{align} 
\tilde \Gamma_f'(\nu)=  - \left( \frac{ \langle  \dot x_f^2 \rangle }{\nu \tilde C_{u\rightarrow f}^{+}(\nu)}\right)''
\\
\tilde \Gamma_f''(\nu)= \frac{K_f}{\nu} -\nu+
 \left( \frac{ \langle  \dot x_f^2 \rangle }{\nu  \tilde C_{u\rightarrow f}^{+}(\nu)}\right)'.
\end{align}
\end{subequations}
From Eq.~\eqref{eq_map2} the mean-squared position and velocity follow as
 \begin{subequations}  \label{eq_map7}
\begin{align}   \label{eq_map7a}
 \langle x_f^2 \rangle =
  \int \frac{ {\rm d}\nu}{2 \pi}  \tilde C^f_{xx}(\nu)
=   \int \frac{ {\rm d}\nu}{2 \pi}   \tilde{f}(\nu) \tilde{f}(-\nu)\tilde C_{xx}(\nu)
\\  \label{eq_map7b}
 \langle  \dot x_f^2 \rangle= 
   \int \frac{ {\rm d}\nu}{2 \pi}  \nu^2 \tilde C^f_{xx}(\nu)
=   \int \frac{ {\rm d}\nu}{2 \pi}  \nu^2 \tilde{f}(\nu) \tilde{f}(-\nu)\tilde C_{xx}(\nu).
 \end{align}
\end{subequations}
The stiffness $K_f$  defined in Eq.~\eqref{K}  follows by dividing   the results in Eq.~\eqref{eq_map7}.
Together, thus, Eqs.~\eqref{eq_map6} and \eqref{eq_map7} allow us to calculate all parameters
of the GLE for the filtered observable, $K_f$, $\Gamma_f(t)$, $\langle \dot x_f^2 \rangle$ and,
using Eq.~\eqref{eq_map3}, also the friction coefficient $\tilde \Gamma_f(0)$
 from the trajectory of the unfiltered observable, the only additional input required is the 
 Fourier transform of the filter function $\tilde{f}(\nu)$.
These formulas   thus determine all parameters of the  time-coarse grained GLE, as will be further  discussed below.

\subsubsection{Extracting unfiltered parameters from  filtered trajectories} 

The other direction is also important for applications, namely reconstructing the parameters
of the GLE for unfiltered observables from filtered trajectories. This is  relevant
when trajectories of a systems are measured using an experimental device that filters the read out, 
which is true for any experimental measurement. The derivation is completely analogous to the one
in the previous section, except that filtered and unfiltered functions are reversed and 
$\tilde{f}(\nu)$ is replaced by $1/\tilde{f}(\nu)$.
 Defining in analogy to Eq.~\eqref{eq_map4}   the function 
\begin{align}  \label{eq_map8}
\tilde C_{f\rightarrow u}(\nu)  = 
\frac{  \imath \nu \tilde C^f_{xx}(\nu)}
{ \tilde{f}(\nu) \tilde{f}(-\nu)   } 
\end{align}
 we obtain 
 \begin{subequations}  \label{eq_map9}
\begin{align} 
\tilde \Gamma'(\nu)=  - \left( \frac{ \langle  \dot x^2 \rangle }{\nu \tilde C_{f\rightarrow u}^{+}(\nu)}\right)''
\\
\tilde \Gamma''(\nu)= \frac{K}{\nu} -\nu+
 \left( \frac{ \langle  \dot x^2 \rangle }{\nu  \tilde C_{f\rightarrow u}^{+}(\nu)}\right)',
\end{align}
\end{subequations}
where $\tilde C_{f\rightarrow u}^{+}(\nu)$ is 
 the Fourier transform of the single-sided time-domain function 
$C^+_{f\rightarrow u}(t) = \theta(t) C_{f\rightarrow u}(t)$
and $C_{f\rightarrow u}(t)$ is the back Fourier transform of $\tilde C_{f\rightarrow u}(\nu)$.
Furthermore, from Eq.~\eqref{eq_map2} we find
 \begin{subequations}  \label{eq_map10}
\begin{align} 
 \langle x^2 \rangle =
  \int \frac{ {\rm d}\nu}{2 \pi}  \tilde C_{xx}(\nu)
=   \int \frac{ {\rm d}\nu}{2 \pi} \frac{\tilde C^f_{xx}(\nu)}{ \tilde{f}(\nu) \tilde{f}(-\nu)}
\\
 \langle  \dot x^2 \rangle= 
   \int \frac{ {\rm d}\nu}{2 \pi}  \nu^2 \tilde C_{xx}(\nu)
=   \int \frac{ {\rm d}\nu}{2 \pi} \frac{ \nu^2 \tilde C^f_{xx}(\nu) }{ \tilde{f}(\nu) \tilde{f}(-\nu)}  .
 \end{align}
\end{subequations}
The potential parameter $K=  \langle  \dot x^2 \rangle/ \langle x^2 \rangle$ for the unfiltered GLE,  defined in analogy to   Eq.~\eqref{K} 
for the filtered observable,  follows from  the results in Eq.~\eqref{eq_map10}.
Together, thus, Eqs.~\eqref{eq_map9} and \eqref{eq_map10} allow to calculate all parameters
of the GLE for the unfiltered observable, $K$, $\Gamma(t)$,  $\langle \dot x^2 \rangle$ and,
using Eq.~\eqref{eq_map3}, also the friction coefficient $\tilde \Gamma(0)$
 from the trajectory of the filtered observable, if the 
 Fourier transform of the filter function $\tilde{f}(\nu)$ is known.

\subsection{Analytical  results for Debye filter}

The mapping between filtered and unfiltered dynamics  derived  so far involves  Fourier transforms of the convolution of the 
 two-point correlation function with the filter function, which for general filter functions
must be done numerically. For the case of a Debye filter,  defined as
\begin{align}  \label{eq_map11}
\tilde{f}(\nu) \tilde{f}(-\nu) = \frac{1}{1+ \lambda^2 \nu^2},
\end{align}
some closed-form results can be obtained by pole analysis.
In the time domain, one possible realization 
 of this filter is
$f(t)=\theta(-t) e^{t/\lambda}/\lambda$ and corresponds to a single-sided normalized  exponential low-pass  filter
with the decay time $\lambda$. 
By residual calculus, the details of which are shown in Appendix \ref{sec_App_DebyeNonMarkov},
the integrals in Eq.~\eqref{eq_map7} can be done with the results
 \begin{subequations}  \label{eq_map12}
\begin{align}  \label{eq_map12a}
 \langle x_f^2 \rangle =  \langle x^2 \rangle  \left[ 
 \frac{1+ \lambda  \tilde \Gamma(-\imath / \lambda)}
 {1+ \lambda  \tilde \Gamma(-\imath / \lambda) + \lambda^2 K} \right],
 \\  \label{eq_map12b}
 \langle  \dot x_f^2 \rangle= 
 \langle \dot x^2 \rangle \left[ 
 \frac{1}
 {1+ \lambda  \tilde \Gamma(-\imath / \lambda) + \lambda^2 K} \right],
 \end{align}
\end{subequations}
where $ \tilde \Gamma(-\imath / \lambda)$ is the kernel Laplace transform, which is a real function. 
It is seen that the mean-squared position decreases due to Debye filtering. 
Also the mean-squared velocity goes down 
due to Debye  filtering for memory functions with a positive Laplace transform, which is the typical scenario. 
Interestingly, the  potential parameter of the filtered observable   
\begin{align} \label{eq_map13}
K_f = \frac{ \langle \dot x_f^2 \rangle}{ \langle x_f^2 \rangle} = 
\frac{K}{ 1+ \lambda  \tilde \Gamma(-\imath / \lambda)}
 \end{align}
goes down for for memory functions with a positive Laplace transform. This  means that the effect
of Debye filtering on  the mean-squared velocity dominates over the filtering effect on the mean-squared position.
If we were dealing with a standard equilibrium system, we would say that 
 the filtering effect on the effective mass, defined by the equipartition theorem 
as $ m^f_{\rm eff}= k_BT  / \langle \dot x_f^2 \rangle$, is more important than the filtering effect 
on the bare harmonic potential strength, which is  given by $ k_BT  / \langle x_f^2 \rangle$.

Combining Eqs.~\eqref{eq_map3}, \eqref{eq_map12}, \eqref{eq_map13} we obtain for the 
Debye-filtered friction coefficient 
\begin{align}  \label{eq_map13b}
\tilde \Gamma_f(0) =  \tilde \Gamma(0) 
 \frac
  {1+ \lambda  \tilde \Gamma(-\imath / \lambda) + \lambda^2 K}
 {(1+ \lambda  \tilde \Gamma(-\imath / \lambda))^2}.
 \end{align}
Interestingly, we see that depending on the values of the memory function  $\tilde \Gamma(-\imath / \lambda)$
and the potential stiffness $K$ of the unfiltered system,  the Debye-filtered friction coefficient  
$\tilde \Gamma_f(0)$ can go up or down compared to the unfiltered friction coefficient  $\tilde \Gamma(0)$.
Only for an unconfined system, i.e. for $K=0$, it is clear that  the Debye-filtered friction coefficient  
$\tilde \Gamma_f(0)$  goes down. 

What do these results mean for the dynamics of the  filtered system? Similar to the overdamped harmonic
oscillator, two  characteristic  time scales are relevant besides the memory time, which we will discuss in the next section:
The persistence time $\tau^f_{\rm per} = 1/\tilde \Gamma_f(0)$, which is the time over which the filtered observable moves ballistically,
and the relaxation time  $\tau^f_{\rm rel} = \tilde \Gamma_f(0)/ K_f$, which measures how quickly the filtered observable relaxes from an excursion. 
From the results for  $\tilde \Gamma_f(0)$ and $ K_f$ in Eqs.~\eqref{eq_map13} and \eqref{eq_map13b} we infer that $\tau^f_{\rm per}$ can increase or decrease due to Debye filtering, but 
the relaxation time $\tau^f_{\rm rel}$ strictly increases due to Debye filtering. This is relevant for coarse-graining procedures, because
it means that the relaxation time of a low-pass filtered coarse-grained observable will go up.

As a side remark, we add that Eqs.~\eqref{eq_map12} and \eqref{eq_map13} can be used as an alternative 
and presumably numerically rather stable method to extract
the Laplace-transformed memory kernel $\tilde \Gamma(-\imath / \lambda)$
from simulation or experimental data by applying a  Debye filter on the  trajectory.

\subsection{Analytical result for the memory function induced by  Debye filtering of Markovian trajectories }

Even for a Debye filter,
 the calculation of  the friction kernel that describes  the filtered observable for a general unfiltered system according
 to  Eq.~\eqref{eq_map6} is analytically prohibitive. 
Here we further simplify the problem by assuming that  the unfiltered trajectory comes
from a Markovian system, i.e., the friction kernel describing the unfiltered observable
 is given by a delta function as 
$\Gamma(t)=2 \gamma \delta(t)$. 
Thus  the response function of the unfiltered system  is according to Eq.~\eqref{eq_response2} given by  
\begin{align}  \label{eq_map14}
\tilde \chi(\nu)  = \left[ K - \nu^2 + \imath \nu \gamma \right]^{-1}. 
\end{align}
Following Eq.~\eqref{eq_Cxx2},  the unfiltered  two-point positional auto-correlation function is given as 
\begin{align}  \label{eq_map15}
&\tilde C_{xx}(\nu)=  \frac{ \langle  \dot x^2 \rangle}{\imath \nu} 
\left[ \tilde \chi(-\nu) - \tilde \chi(\nu)  \right].
\end{align}
Combining Eqs.~\eqref{eq_response2},   \eqref{eq_Cxx2},  \eqref{eq_map2}, \eqref{eq_map11},  \eqref{eq_map14}
and  \eqref{eq_map15} we arrive at the implicit  equation that determines the parameters of the 
Debye-filtered GLE, namely $\langle  \dot x_f^2 \rangle$, $K_f$ and 
$ \tilde \Gamma_f(\nu)$, in terms of the parameters of the 
unfiltered GLE, namely $\langle  \dot x^2 \rangle$, $K$ and  $\gamma$, 
which reads
\begin{align}  \label{eq_map16}
& \langle  \dot x_f^2 \rangle \left[ \frac{1}{K_f - \nu^2 - \imath \nu \tilde \Gamma_f(-\nu)}
- \frac{1}{K_f - \nu^2 + \imath \nu \tilde \Gamma_f(\nu)} \right] = \nonumber \\
& \frac{\langle  \dot x^2 \rangle }{1+ \lambda^2 \nu^2}
\left[ \frac{1}{K - \nu^2 - \imath \nu  \gamma}
- \frac{1}{K- \nu^2 + \imath \nu \gamma} \right] .
\end{align}
Note that the filtering mixes anticausal and causal poles on the right-hand side of the equation. 
Separating the 6 poles of the equation into terms that  are purely causal and anticausal,
 the details of the calculation are shown in Appendix \ref{sec_App_DebyeMarkov},
 yields  the final results
 \begin{subequations}   \label{eq_map17}
\begin{align} \label{eq_map16a}
\frac{ \langle x_f^2 \rangle}{ \langle x^2 \rangle } =  
 \frac{1+ \lambda  \gamma}
 {1+ \lambda \gamma + \lambda^2 K}
 \\ \label{eq_map16b}
 \frac{ \langle  \dot x_f^2 \rangle}{  \langle \dot x^2 \rangle} = 
 \frac{1}
 {1+ \lambda \gamma + \lambda^2 K} 
 \\ \label{eq_map16c}
K_f = \frac{ \langle \dot x_f^2 \rangle}{ \langle x_f^2 \rangle} = 
\frac{K}{ 1+ \lambda  \gamma},
 \end{align}
\end{subequations}
which agree with the results in the preceding section for a general non-Markovian unfiltered system
Eqs.~\eqref{eq_map12} and  \eqref{eq_map13}.
It is seen that both mean-squared position and mean-squared velocity go down 
due to Debye-filtering in a way so  that the filtered potential parameter 
$K_f$  goes also  down.
The closed-form  result for the memory kernel of the filtered variable reads
\begin{align}  \label{eq_map18}
\tilde \Gamma_f(\nu)  =  
 \frac{  \gamma_f}
 {1+ \imath \tau_f \nu},
  \end{align}
which means that Debye-filtering induces an exponentially decaying memory kernel,
which in the time domain reads 
\begin{align}  \label{eq_map18b}
\Gamma_f(t)= \frac{ \gamma_f }{\tau_f}   e^{-t/\tau_f}, 
  \end{align}
with the effective friction
and memory time given by 
  \begin{subequations}   \label{eq_map19}
\begin{align} \label{eq_map19a}
\gamma_f &= 
 \frac {\gamma(1+ \lambda \gamma + \lambda^2 K)}
 {(1+ \lambda  \gamma)^2}, \\  \label{eq_map19b}
 \tau_f &=  \frac {\lambda}
 {1+ \lambda  \gamma}.
  \end{align}
\end{subequations}
We remark that the result for the friction coefficient in  Eq.~\eqref{eq_map19a} 
is consistent  with the  general result in Eq.~\eqref{eq_map13b}.
We see that Debye filtering of a Markovian trajectory  
 produces non-Markovianity in the form of a single-exponential memory kernel 
 with a memory time $\tau_f$ that is strictly shorter than the Debye filter decay time $\lambda$. 
The friction coefficient $\gamma_f$ of the filtered trajectory can be larger or smaller
than the friction coefficient $\gamma$ of the unfiltered  Markovian trajectory,
only for an unconfined trajectory with $K=0$ do we know for sure that the 
friction coefficient $\gamma_f$ of the filtered trajectory goes down compared to 
the friction coefficient $\gamma$ of the unfiltered  Markovian trajectory,

\section{Discussion}

We have shown that the dynamics of convolution filtered observables obey the Liouville equation
just as regular observables do.
Based on this finding we derived the GLE for filtered observables by exact projection in phase space, which has the 
same structure as the GLE for regular observables. 
We derived explicit transformation formulas that allow us to calculate the parameters of the 
filtered GLE from parameters of the unfiltered GLE and vice versa. 

There are two major  applications of our filter-projection approach:
Low-pass filtering eliminates fast data components and thereby yields a temporally coarse-grained model.
Our filter-projection approach not only shows that the GLE is the exact equation of motion for such 
temporally  coarse-grained variable, it also provides  all the parameters of the filtered GLE. 

Conversely, the elimination of slow or periodic data components by high-pass or band-stop filtering is important 
in many practical situations. 
The standard  approach towards
such systems would be to use non-equilibrium statistical mechanics methods, for example based on a time-dependent Hamiltonian.
 The concept of a time-dependent
Hamiltonian derives from splitting the autonomous time-independent Hamiltonian, which encompasses the system components that  cause 
the slow transient  or seasonal  dynamics, 
into the system of interest  and its slowly evolving  environment. The coupling between the system of interest
and the environment then leads to time-dependent terms in the system Hamiltonian.  
Recently,  
time-dependent projection techniques have been used  to derive  non-equilibrium GLEs 
for systems that are described by time-dependent Liouville operators  \cite{meyer_non-stationary_2017,Vrugt2019}.
Even more recently it was shown that systems described by time-dependent Hamiltonians can also be treated
by time-independent projection techniques  \cite{Netz2024}. 
In the present filter-projection approach there is no need to introduce a time-dependent Hamiltonian or Liouville operator,
rather, the Hamiltonian can be considered time-independent and 
 large enough so that it includes  the system components causing the  slow and transient  dynamics.
The slow and transient data components  are then removed by filtering. In that sense,
our filter-projection approach constitutes an alternative to the usual non-equilibrium approach to systems 
that exhibit slow transient effects, such as active systems, weather or financial data.

In this work we assume that the trajectory of a given observable
 comes from simulations or experiments and obeys classical autonomous Hamiltonian dynamics, the interesting case of quantum
 dynamics has  not been treated. 
Also,  the question of the existence of an optimal  observable  for characterizing the system dynamics  is important but was  not considered. 
Finally, we remark that although  the effect of discretization of continuous data can not be described as 
a convolution filter,   the effect of discretization is  expected to be reduced by prior  low-pass filtering. 

\begin{acknowledgments}
We acknowledge support by Deutsche Forschungsgemeinschaft Grant CRC 1114 "Scaling Cascades in Complex System", 
Project 235221301, Project B03 
and by the ERC Advanced Grant 835117 NoMaMemo
and by the Infosys Foundation. 
 We gratefully acknowledge  computing time on the HPC clusters at the physics department and ZEDAT, FU Berlin.
\end{acknowledgments}

\appendix

\begin{widetext}

\section{Derivation of the GLE for filtered observable \label{sec_App_GLE}}

We begin with the derivation of a few important properties of the Mori projection operator in Eq. \eqref{eq_mori_projection},
which we split into  three parts according to 
\begin{align}
\label{app_mori1}
 {\cal P}  A(\omega,t) = {\cal P}_1  A(\omega,t) +{\cal P}_2 A(\omega,t) +{\cal P}_3 A(\omega,t)
   \end{align}
with
 \begin{align}
{\cal P}_1  A(\omega,t) =
 \langle   A(\omega,t)  \rangle,
   \end{align}
 \begin{align}
{\cal P}_2  A(\omega,t) =
\frac{\langle   A(\omega,t)  {\cal L} B_f (\omega, t_P)  \rangle}  {\langle ( {\cal L}B_f(\omega, t_P) )^2 \rangle} 
  {\cal L}B_f (\omega, t_P),
    \end{align}
 \begin{align}
{\cal P}_3  A(\omega,t) =
\frac{\langle   A(\omega,t)  (B_f (\omega, t_P) - \langle B_f \rangle )\rangle}  {\langle (B_f (\omega, t_P) -  \langle B_f \rangle  )^2 \rangle} 
 (B_f (\omega, t_P) - \langle B_f \rangle ).
   \end{align}

The linearity of the Mori projection,
 i.e., the fact that for two arbitrary observables $A(\omega,t)$ and $C(\omega,t')$ the property 
 ${\cal P} (c_1 A(\omega,t) + c_2 C(\omega,t'))=c_1 {\cal P} A(\omega,t)  + c_2 {\cal P} C(\omega,t')$ holds,
 is self-evident, ${\cal Q}$ thereby  also follows to be linear. 
The idempotency of  ${\cal P}$, i.e., the fact that  ${\cal P}^2= {\cal P}$, is not  so self-evident and the proof will be  split  into three parts. First,
 \begin{align} \label{app_mori2}
{\cal P} {\cal P}_1  A(\omega,t) & =
 \langle     A(\omega,t)  \rangle  \left[ 1 +
    \frac{\langle    {\cal L} B_f (\omega, t_P)  \rangle}  {\langle ( {\cal L}B_f(\omega, t_P) )^2 \rangle} 
  {\cal L}B_f (\omega, t_P) +
 \frac{ \langle B_f (\omega, t_P) - \langle B_f \rangle \rangle}  {\langle (B_f (\omega, t_P) -  \langle B_f \rangle  )^2 \rangle} 
 (B_f (\omega, t_P) - \langle B_f \rangle ) \right]
 \nonumber \\
 & =   {\cal P}_1    A(\omega,t).
   \end{align}
  We used that $ \langle   \langle  A(\omega,t)  \rangle  \rangle =  \langle  A(\omega,t)  \rangle$, which holds since the 
probability distribution $\rho_P(\omega)$  in Eq.  \eqref{projectdist} is normalized. 
For the second term we  used the anti-self-adjointedness of the Liouville operator  $ {\cal L}$ and the stationarity of the 
projection distribution, i.e.  $ {\cal L}\rho_P(\omega)=0$.
Second,
 \begin{align} \label{app_mori3}
{\cal P} {\cal P}_2  A(\omega,t) & =
\frac{\langle   A(\omega,t)  {\cal L} B_f (\omega, t_P)  \rangle}  {\langle ( {\cal L}B_f(\omega, t_P) )^2 \rangle}    \nonumber  \\
 & \times   \left[ \langle   {\cal L}B_f (\omega, t_P) \rangle   +
    \frac{\langle  (   {\cal L}B_f (\omega, t_P))^2   \rangle}  {\langle ( {\cal L}  B_f(\omega, t_P) )^2 \rangle} 
  {\cal L}B_f (\omega, t_P) +
 \frac{  \langle (B_f (\omega, t_P) - \langle B_f \rangle )  {\cal L}B_f (\omega, t_P)\rangle }  {\langle (B_f (\omega, t_P) -  \langle B_f \rangle  )^2 \rangle} 
 (B_f (\omega, t_P) - \langle B_f \rangle ) \right]   \nonumber \\
 & =   {\cal P}_2   B(\omega,t)  ,
   \end{align}
where again we used  the anti-self-adjointedness of  $ {\cal L}$ and the stationarity of   $\rho_P(\omega)$.
Third,
 \begin{align} \label{app_mori4}
& {\cal P} {\cal P}_3  A(\omega,t)  =
\frac{\langle   A(\omega,t)  (B_f (\omega, t_P) - \langle B_f \rangle )\rangle}  {\langle (B_f (\omega, t_P) -  \langle B_f \rangle  )^2 \rangle} 
  \nonumber \\
 & \times   \left[ \langle B_f (\omega, t_P) - \langle B_f \rangle  \rangle   +
    \frac{\langle  (B_f (\omega, t_P) - \langle B_f \rangle )   {\cal L}B_f (\omega, t_P)   \rangle}  {\langle ( {\cal L}  B_f(\omega, t_P) )^2 \rangle} 
  {\cal L}B_f (\omega, t_P) +
 \frac{  \langle (B_f (\omega, t_P) - \langle B_f \rangle )^2 \rangle }  {\langle (B_f (\omega, t_P) -  \langle B_f \rangle  )^2 \rangle} 
 (B_f (\omega, t_P) - \langle B_f \rangle ) \right]   \nonumber  \\
 & =   {\cal P}_3   B(\omega,t)  .
   \end{align}
Adding Eqs.  \eqref{app_mori2},  \eqref{app_mori3},  \eqref{app_mori4} we see
that ${\cal P}^2= {\cal P}( {\cal P}_1 +{\cal P}_2 + {\cal P}_3)= {\cal P}_1 +{\cal P}_2 + {\cal P}_3= {\cal P}$
and thus 
${\cal P}$ is idempotent. From the idempotency of ${\cal P}$  the idempotency of  ${\cal Q}$  also follows, this can be easily seen from 
\begin{align}
{\cal Q}^2 A(\omega,t)= (1- {\cal P})^2 A(\omega,t) = (1-2 {\cal P} + {\cal P}^2) A(\omega,t) = 
(1-{\cal P}) A(\omega,t) ={\cal Q} A(\omega,t).
\end{align} 
The self-adjointedness of ${\cal P}$, Eq. \eqref{eq_projection_orthogonal}, is straightforwardly proven by writing
\begin{align}
&\langle C(\omega,t) {\cal P} A(\omega,t')  \rangle = \\
& \langle C(\omega,t) \rangle \langle A(\omega,t')  \rangle
+    \langle C(\omega,t)   {\cal L} B_f(\omega,t_P)  \rangle
   \frac{ \langle  A(\omega,t')  {\cal L} B_f(\omega,t_P) \rangle}  
   {\langle  ( {\cal L} B_f(\omega,t_P) )^2 \rangle} 
 +    \langle C(\omega,t) ( B_f (\omega,t_P)-\langle  B_f \rangle )  \rangle
\frac{   \langle A(\omega,t')   (B_f (\omega,t_P)-\langle  B_f \rangle )  \rangle }  
 {\langle (B_f (\omega,t_P) - \langle  B_f \rangle )^2 \rangle} \\
&= \langle A(\omega,t') {\cal P} C(\omega,t)   \rangle.
\end{align} 
By using ${\cal Q}=1-{\cal P}$ we see straightforwardly that ${\cal Q}$ is also self-adjoint.

Using similar arguments as above, one can show that  ${\cal P} c$= c, 
  ${\cal P}  (B_f (\omega,t_P) - \langle  B_f \rangle )=  B_f (\omega,t_P) - \langle  B_f \rangle$,
   ${\cal P}    {\cal L} B_f(\omega,t_P) =    {\cal L} B_f(\omega, t_P $,
   from which follows that also   ${\cal P}  B_f (\omega, t_P) =  B_f (\omega, t_P)   $.
   From these relations  we can directly conclude that 
   ${\cal Q} c$= 0, 
  ${\cal Q}  (B_f (\omega,t_P) - \langle  B_f \rangle )=  0$,
   ${\cal Q}   {\cal L} B_f(\omega,t_P)  =   0$,
   and also   ${\cal Q}  B_f(\omega,t_P)   =  0 $.
   From the idempotency of ${\cal P}$ or ${\cal Q}$ it  follows that 
   ${\cal P} {\cal Q}= {\cal P} (1-{\cal P})=  {\cal P} -{\cal P}^2= 0$
   and, similarly,  ${\cal Q} {\cal P}=0$, thus, the operators  ${\cal P}$ and $ {\cal Q}$ are orthogonal to 
   each other.

We now derive the GLE and  consider the first term on the right-hand side
 in Eq. \eqref{eq_GLE3}, which, apart from the propagator in front, 
splits into three terms
\begin{align}
\label{app_GLE1}
  {\cal P}   {\cal L}^2   B_f(\omega,t_P) =  ( {\cal P}_1 +{\cal P}_2 + {\cal P}_3) {\cal L}^2   B_f(\omega,t_P).
\end{align} 
The first term  is given by 
\begin{align}
\label{app_GLE2}
  {\cal P}_1  {\cal L}^2   B_f(\omega,t_P)=
\langle  {\cal L}^2   B_f(\omega,t_P) \rangle = 0,
\end{align} 
where we used the anti-self-adjointedness of  $ {\cal L}$ and the stationarity of   $\rho_P(\omega)$.
The second term  reads
\begin{align}
\label{app_GLE3}
  {\cal P}_2  {\cal L}^2   B_f(\omega,t_P) & =
 \frac{  \langle  (   {\cal L}   B_f(\omega,t_P))   {\cal L}^2   B_f(\omega,t_P)  \rangle}
   {\langle  ( {\cal L} B_f(\omega,t_P) )^2 \rangle}  {\cal L}  B_f(\omega,t_P)  =  0.
\end{align} 
 The third term reads 
\begin{align}
\label{app_GLE4}
 {\cal P}_3  {\cal L}^2   B_f(\omega,t_P) & =
 \frac{  \langle  (B_f (\omega, t_P) - \langle B_f \rangle )   {\cal L}^2   B_f(\omega,t_P)  \rangle }
 {\langle (B_f (\omega,t_P) - \langle  B_f \rangle )^2 \rangle} (B_f (\omega, t_P) - \langle B_f \rangle )  
\nonumber \\
&=  -  \frac{  (   {\cal L}   B_f(\omega,t_P))^2  \rangle }
 {\langle (B_f (\omega,t_P) - \langle  B_f \rangle )^2 \rangle} (B_f (\omega, t_P) - \langle B_f \rangle ) .
\end{align}

Combining the results in Eqs. \eqref{app_GLE1},  \eqref{app_GLE2},  \eqref{app_GLE3},  \eqref{app_GLE4}, 
the first term in Eq. \eqref{eq_GLE3} reads
\begin{align}
\label{app_GLE8}
e^{(t-t_P)   {\cal L}}          {\cal P}  {\cal L}^2  B_f(\omega,t_P)  &= 
- K_f e^{(t-t_P)  {\cal L}}    (B_f (\omega, t_P) - \langle B_f \rangle ) \nonumber \\
&= - K_f  (B_f (\omega, t) - \langle B_f \rangle )
\end{align} 
where $K_f$ is  defined in Eqs.  \eqref{GLEK}.

We now consider the last term on the right-hand side  in Eq. \eqref{eq_GLE3}, which 
 reads, without the time integral and the propagator in front, 
\begin{align}
\label{app_GLE8b}
  {\cal P}   {\cal L}   F(\omega, s + t_P) =  ( {\cal P}_1 +{\cal P}_2 + {\cal P}_3) {\cal L}   F(\omega, s + t_P).
\end{align} 
The first term  is given by 
\begin{align}
\label{app_GLE9}
 {\cal P}_1  {\cal L}   F(\omega, s + t_P) & = \langle  {\cal L}    F(\omega,s+t_P)  \rangle  \rangle = 0,
\end{align} 
where we used the anti-self-adjointedness of  $ {\cal L}$ and the stationarity of   $\rho_P(\omega)$.
The second term reads
\begin{align}
\label{app_GLE10}
 {\cal P}_2  {\cal L}   F(\omega, s + t_P)  & =
 \frac{  \langle   ( {\cal L}   F(\omega, s + t_P) )    {\cal L}   B_f(\omega,t_P)  \rangle}
   {\langle  ( {\cal L} B_f(\omega,t_P) )^2 \rangle}  {\cal L}  B_f(\omega,t_P) 
   = -   \frac{  \langle     F(\omega, s + t_P)     {\cal L}^2   B_f(\omega,t_P)  \rangle}
   {\langle  ( {\cal L} B_f(\omega,t_P) )^2 \rangle}  {\cal L}  B_f(\omega,t_P) \nonumber \\
&=  -  \frac{   \langle  ( {\cal Q}    F(\omega, s + t_P)  )    {\cal L}^2   B_f(\omega,t_P)  \rangle}
   {\langle  ( {\cal L} B_f(\omega,t_P) )^2 \rangle}  {\cal L}  B_f(\omega,t_P) 
=  - \frac{  \langle     F(\omega, s + t_P)    {\cal Q}  {\cal L}^2   B_f(\omega,t_P)  \rangle}
   {\langle  ( {\cal L} B_f(\omega,t_P) )^2 \rangle} {\cal L}   B_f(\omega,t_P) \nonumber \\
&=  - \frac{  \langle     F(\omega, s + t_P)   F(\omega, t_P)   \rangle}
   {\langle  ( {\cal L} B_f(\omega,t_P) )^2 \rangle} {\cal L}   B_f(\omega,t_P)
   =  - \frac{  \langle     F(\omega, s )   F(\omega, 0)   \rangle}
   {\langle  ( {\cal L} B_f(\omega,t_P) )^2 \rangle} {\cal L}   B_f(\omega,t_P),
\end{align} 
where we used the anti-self-adjointedness of  $ {\cal L}$ and the stationarity of   $\rho_P(\omega)$,
the idempotency and self-adjointedness of  ${\cal Q}$ as well as 
 the definition of the complementary force in Eq. \eqref{eq_F_operator}.
The third term reads
\begin{align}
\label{app_GLE11}
 {\cal P}_3  {\cal L}   F(\omega, s + t_P)  & =
 \frac{  \langle   ( {\cal L}   F(\omega, s + t_P) )   (B_f (\omega,t_P) - \langle  B_f \rangle )  \rangle}
 {\langle (B_f (\omega,t_P) - \langle  B_f \rangle )^2 \rangle} (B_f (\omega, t_P) - \langle B_f \rangle )   \nonumber \\
& = -
 \frac{  \langle     F(\omega, s + t_P)   {\cal L}    B_f (\omega,t_P)  \rangle}
 {\langle (B_f (\omega,t_P) - \langle  B_f \rangle )^2 \rangle} (B_f (\omega, t_P) - \langle B_f \rangle )   \nonumber \\
& = -  \frac{  \langle   ({\cal Q}   F(\omega, s + t_P) )    {\cal L}    B_f (\omega,t_P)  \rangle}
 {\langle (B_f (\omega,t_P) - \langle  B_f \rangle )^2 \rangle} (B_f (\omega, t_P) - \langle B_f \rangle )   \nonumber \\
& = -  \frac{  \langle    F(\omega, s + t_P)   {\cal Q}   {\cal L}    B_f (\omega,t_P)  \rangle}
 {\langle (B_f (\omega,t_P) - \langle  B_f \rangle )^2 \rangle} (B_f (\omega, t_P) - \langle B_f \rangle )  =0,
 \end{align} 
where we used the anti-self-adjointedness of  $ {\cal L}$ and the stationarity of   $\rho_P(\omega)$, 
 the idempotency and self-adjointedness of  ${\cal Q}$,
 the definition of the complementary force in Eq. \eqref{eq_F_operator}
 as well as  the fact that $ {\cal Q}   {\cal L}    B_f (\omega,t_P) =0$. 
Combining the results in Eqs. \eqref{app_GLE8b},  \eqref{app_GLE9},  \eqref{app_GLE10},  \eqref{app_GLE11}, 
the integrand in the last term on the right-hand side  in Eq. \eqref{eq_GLE3} reads
\begin{align}
\label{app_GLE12}
e^{(t-t_P-s)   {\cal L}}          {\cal P}  {\cal L}   F(\omega, s + t_P)  &= 
- \Gamma_f(s)  e^{(t-t_P-s)  {\cal L}}    {\cal L}  B_f (\omega, t_P)  \nonumber \\
&= - \Gamma_f(s)   {\cal L}  B_f (\omega, t-s) 
\end{align} 
where $\Gamma_f(s) $ is  defined in Eqs.  \eqref{eq_mori_memory}.
Inserting the results in Eqs. \eqref{app_GLE8} and  \eqref{app_GLE12} into
 the general GLE in Eq. \eqref{eq_GLE3} we obtain  the explicit Mori GLE  in Eq.  \eqref{eq_mori_GLE}.

%\section{Derivation of time-homogeneous GLE \label{sec_App_GLEx}}
\section{Derivation of the Debye-filtered mean-squared position and  mean-squared  velocity \label{sec_App_DebyeNonMarkov}}

Here we derive Eqs. \eqref{eq_map12a} and  \eqref{eq_map12b} by residual calculus. 
Combining Eqs. \eqref{eq_map7a} and   \eqref{eq_map15}  we obtain 
\begin{align}   \label{AppB1}
 \langle x_f^2 \rangle =
  \int \frac{ {\rm d}\nu}{2 \pi}  \tilde C^f_{xx}(\nu) =   \int \frac{ {\rm d}\nu}{2 \pi}  \nu^2 \tilde{f}(\nu) \tilde{f}(-\nu)\tilde C_{xx}(\nu)
=   \int \frac{ {\rm d}\nu}{2 \pi}   \tilde{f}(\nu) \tilde{f}(-\nu)
\frac{ \langle  \dot x^2 \rangle}{\imath \nu} 
\left[ \tilde \chi(-\nu) - \tilde \chi(\nu)  \right].
\end{align}
Inserting the Fourier transform of the Debye filter from Eq.  \eqref{eq_map11} we obtain 
\begin{align}   \label{AppB2}
 \langle x_f^2 \rangle =
 \langle  \dot x^2 \rangle
   \int \frac{ {\rm d}\nu}{2 \pi}   
\frac{      1      }{\imath \nu (1+ \lambda^2 \nu^2) } 
\left[ \tilde \chi(-\nu) - \tilde \chi(\nu)  \right].
\end{align}
The response function of the unfiltered system $\tilde \chi(\nu) $  is given in  Eq. \eqref{eq_map14}. 
Since the time-domain response function $\chi(t)$  is a single-sided decaying function,
$\tilde \chi(\nu) $ has no poles in the lower complex plane while 
$\tilde \chi(-\nu) $ has no poles in the upper complex plane. We thus close the integration contour in Eq. \eqref{AppB2}
of the integrand proportional to $\tilde \chi(\nu)$ in the lower complex plane and of  the integrand proportional to $\tilde \chi(-\nu)$ in the lower complex plane.
The residuals of the  three poles at $\nu=0$, $\nu=\pm \imath /\lambda$ give rise to the final result 
\begin{align}   \label{AppB3}
 \langle x_f^2 \rangle =
 \langle  \dot x^2 \rangle
 \left[ \frac{1}{K} - \frac{\lambda^2}{ 1+ K \lambda^2 +  \lambda  \tilde \Gamma(-\imath/\lambda)} \right].
\end{align}
Similarly, by combining Eqs. \eqref{eq_map7b}, \eqref{eq_map2} and  \eqref{eq_map15} we obtain 
\begin{align}   \label{AppB4}
 \langle \dot x_f^2 \rangle =
 \langle  \dot x^2 \rangle
   \int \frac{ {\rm d}\nu}{2 \pi}   
\frac{      \nu^2      }{\imath \nu (1+ \lambda^2 \nu^2) } 
\left[ \tilde \chi(-\nu) - \tilde \chi(\nu)  \right],
\end{align}
which  by residual calculus yields the final result 
\begin{align}   \label{AppB5}
 \langle \dot x_f^2 \rangle =
\frac{  \langle  \dot x^2 \rangle  }{ 1+ K \lambda^2 +  \lambda  \tilde \Gamma(-\imath/\lambda)}
\end{align}
in Eq. \eqref{eq_map12b}.
 Eq. \eqref{eq_map12a} follows from  Eq.  \eqref{AppB3}  by dividing by $K=   \langle \dot x^2 \rangle /  \langle  x^2 \rangle$.

\section{Derivation of  Debye-filtered memory kernel for Markovian unfiltered system \label{sec_App_DebyeMarkov}}

We start the derivation of Eqs.  \eqref{eq_map18}, \eqref{eq_map19a}  and \eqref{eq_map19b} by  slightly rewriting Eq.    \eqref{eq_map16} as

\begin{align}  \label{AppC1}
&  \tilde \chi_f(-\nu) - \tilde \chi_f(\nu)  = 
 \frac{1}{K_f - \nu^2 - \imath \nu \tilde \Gamma_f(-\nu)}
- \frac{1}{K_f - \nu^2 + \imath \nu \tilde \Gamma_f(\nu)} \nonumber \\
&=  \frac{\langle  \dot x^2 \rangle }{ \langle  \dot x_f^2 \rangle}
\left[ \frac{1}{K - \nu^2 - \imath \nu  \gamma}
- \frac{1}{K- \nu^2 + \imath \nu \gamma} \right]
\frac{1} {1+ \lambda^2 \nu^2}
 \nonumber \\
&=  \frac{\langle  \dot x^2 \rangle }{ \langle  \dot x_f^2 \rangle}
\left[ \frac{1}{\nu^2 - \imath \nu  \gamma - K   }
- \frac{1}{\nu^2 + \imath \nu \gamma - K } \right] 
\frac{\lambda^{-2}} { \nu^2 + \lambda^{-2} }  \nonumber \\
&=  \frac{\langle  \dot x^2 \rangle }{ \langle  \dot x_f^2 \rangle}
\left[ \frac{1} {(\nu - \nu_1) (\nu - \nu_2)}
- \frac{1} {(\nu - \nu_3) (\nu - \nu_4)}  \right] 
\frac{\lambda^{-2}} {(\nu - \nu_5) (\nu - \nu_6)}.
\end{align}
As already mentioned in the main text,
while on the left-hand side of Eq.  \eqref{AppC1}  the expressions $ \tilde \chi_f(-\nu)$ and $ \tilde \chi_f(\nu)$ correspond
to separate anticausal and causal terms,  
 the filtering mixes the six anticausal and causal poles on the right-hand side of Eq.  \eqref{AppC1}.
The six poles are given by 
\begin{align}  \label{AppC2}
\nu_{1,2} = \frac{\imath \gamma}{2} \pm \sqrt{K-\gamma^2/4}, \nonumber \\
\nu_{3,4} = -  \frac{\imath \gamma}{2} \pm \sqrt{K-\gamma^2/4}, \nonumber \\
\nu_{5,6} =  \pm \imath \lambda^{-1}.
\end{align}
The poles $ \nu_{1,2}$ are located in the upper complex plane and are thus causal,
the poles $ \nu_{3,4}$ are located in the lower complex plane and are thus anticausal,
the pole $ \nu_{5}= \imath \lambda^{-1}$ is  causal and
the pole $ \nu_{6}= - \imath \lambda^{-1}$ is  anticausal.
By basic algebraic operations the right-hand side of Eq.  \eqref{AppC1} can be  separated into a causal 
and an anticausal part. The causal part turns out to be given by 
\begin{align}  \label{AppC3}
 \tilde \chi_f(\nu)  =   \frac{1}{K_f - \nu^2 + \imath \nu \tilde \Gamma_f(\nu)} 
 =\frac{ \imath \gamma + \imath \lambda^{-1} - \nu}
 {(\nu - \nu_1) (\nu - \nu_2)(\nu - \nu_5)}.
\end{align}
Solving for $ \tilde \Gamma_f(\nu)$ one obtains the result in Eq.  \eqref{eq_map18}
with the  friction coefficient $\gamma_f$ given by Eq. \eqref{eq_map19a}  and
the  memory time $\tau_f$ given by Eq.  \eqref{eq_map19b}.

\end{widetext}

\bibliography{filterGLE}

\begin{thebibliography}{57}%
\makeatletter
\providecommand \@ifxundefined [1]{%
 \@ifx{#1\undefined}
}%
\providecommand \@ifnum [1]{%
 \ifnum #1\expandafter \@firstoftwo
 \else \expandafter \@secondoftwo
 \fi
}%
\providecommand \@ifx [1]{%
 \ifx #1\expandafter \@firstoftwo
 \else \expandafter \@secondoftwo
 \fi
}%
\providecommand \natexlab [1]{#1}%
\providecommand \enquote  [1]{``#1''}%
\providecommand \bibnamefont  [1]{#1}%
\providecommand \bibfnamefont [1]{#1}%
\providecommand \citenamefont [1]{#1}%
\providecommand \href@noop [0]{\@secondoftwo}%
\providecommand \href [0]{\begingroup \@sanitize@url \@href}%
\providecommand \@href[1]{\@@startlink{#1}\@@href}%
\providecommand \@@href[1]{\endgroup#1\@@endlink}%
\providecommand \@sanitize@url [0]{\catcode `\\12\catcode `\$12\catcode
  `\&12\catcode `\#12\catcode `\^12\catcode `\_12\catcode `\%12\relax}%
\providecommand \@@startlink[1]{}%
\providecommand \@@endlink[0]{}%
\providecommand \url  [0]{\begingroup\@sanitize@url \@url }%
\providecommand \@url [1]{\endgroup\@href {#1}{\urlprefix }}%
\providecommand \urlprefix  [0]{URL }%
\providecommand \Eprint [0]{\href }%
\providecommand \doibase [0]{https://doi.org/}%
\providecommand \selectlanguage [0]{\@gobble}%
\providecommand \bibinfo  [0]{\@secondoftwo}%
\providecommand \bibfield  [0]{\@secondoftwo}%
\providecommand \translation [1]{[#1]}%
\providecommand \BibitemOpen [0]{}%
\providecommand \bibitemStop [0]{}%
\providecommand \bibitemNoStop [0]{.\EOS\space}%
\providecommand \EOS [0]{\spacefactor3000\relax}%
\providecommand \BibitemShut  [1]{\csname bibitem#1\endcsname}%
\let\auto@bib@innerbib\@empty
%</preamble>
\bibitem [{\citenamefont {Levitt}\ and\ \citenamefont
  {Warshel}(1975)}]{Levitt1975}%
  \BibitemOpen
  \bibfield  {author} {\bibinfo {author} {\bibfnamefont {M.}~\bibnamefont
  {Levitt}}\ and\ \bibinfo {author} {\bibfnamefont {A.}~\bibnamefont
  {Warshel}},\ }\bibfield  {title} {\bibinfo {title} {{Computer simulation of
  protein folding}},\ }\href@noop {} {\bibfield  {journal} {\bibinfo  {journal}
  {Nature}\ }\textbf {\bibinfo {volume} {253}},\ \bibinfo {pages} {694}
  (\bibinfo {year} {1975})}\BibitemShut {NoStop}%
\bibitem [{\citenamefont {Kmiecik}\ \emph {et~al.}(2016)\citenamefont
  {Kmiecik}, \citenamefont {Gront}, \citenamefont {Kolinski}, \citenamefont
  {Wieteska}, \citenamefont {Dawid},\ and\ \citenamefont
  {Kolinski}}]{Kmiecik2016}%
  \BibitemOpen
  \bibfield  {author} {\bibinfo {author} {\bibfnamefont {S.}~\bibnamefont
  {Kmiecik}}, \bibinfo {author} {\bibfnamefont {D.}~\bibnamefont {Gront}},
  \bibinfo {author} {\bibfnamefont {M.}~\bibnamefont {Kolinski}}, \bibinfo
  {author} {\bibfnamefont {L.}~\bibnamefont {Wieteska}}, \bibinfo {author}
  {\bibfnamefont {A.~E.}\ \bibnamefont {Dawid}},\ and\ \bibinfo {author}
  {\bibfnamefont {A.}~\bibnamefont {Kolinski}},\ }\bibfield  {title} {\bibinfo
  {title} {{Coarse-Grained Protein Models and Their Applications}},\
  }\href@noop {} {\bibfield  {journal} {\bibinfo  {journal} {Chem. Rev.}\
  }\textbf {\bibinfo {volume} {116}},\ \bibinfo {pages} {7898} (\bibinfo {year}
  {2016})}\BibitemShut {NoStop}%
\bibitem [{\citenamefont {Walmod}\ \emph {et~al.}(2004)\citenamefont {Walmod},
  \citenamefont {Hartmann-Petersen}, \citenamefont {Prag}, \citenamefont
  {Lepekhin}, \citenamefont {Röpke}, \citenamefont {Berezin},\ and\
  \citenamefont {Bock}}]{walmod2004}%
  \BibitemOpen
  \bibfield  {author} {\bibinfo {author} {\bibfnamefont {P.~S.}\ \bibnamefont
  {Walmod}}, \bibinfo {author} {\bibfnamefont {R.}~\bibnamefont
  {Hartmann-Petersen}}, \bibinfo {author} {\bibfnamefont {S.}~\bibnamefont
  {Prag}}, \bibinfo {author} {\bibfnamefont {E.~L.}\ \bibnamefont {Lepekhin}},
  \bibinfo {author} {\bibfnamefont {C.}~\bibnamefont {Röpke}}, \bibinfo
  {author} {\bibfnamefont {V.}~\bibnamefont {Berezin}},\ and\ \bibinfo {author}
  {\bibfnamefont {E.}~\bibnamefont {Bock}},\ }\bibfield  {title} {\bibinfo
  {title} {{Cell-cycle-dependent regulation of cell motility and determination
  of the role of Rac1}},\ }\href@noop {} {\bibfield  {journal} {\bibinfo
  {journal} {Experimental Cell Research}\ }\textbf {\bibinfo {volume} {295}},\
  \bibinfo {pages} {407} (\bibinfo {year} {2004})}\BibitemShut {NoStop}%
\bibitem [{\citenamefont {Franzke}\ \emph {et~al.}(2015)\citenamefont
  {Franzke}, \citenamefont {O'Kane}, \citenamefont {Berner}, \citenamefont
  {Williams},\ and\ \citenamefont {Lucarini}}]{franzke2015stochastic}%
  \BibitemOpen
  \bibfield  {author} {\bibinfo {author} {\bibfnamefont {C.~L.~E.}\
  \bibnamefont {Franzke}}, \bibinfo {author} {\bibfnamefont {T.~J.}\
  \bibnamefont {O'Kane}}, \bibinfo {author} {\bibfnamefont {J.}~\bibnamefont
  {Berner}}, \bibinfo {author} {\bibfnamefont {P.~D.}\ \bibnamefont
  {Williams}},\ and\ \bibinfo {author} {\bibfnamefont {V.}~\bibnamefont
  {Lucarini}},\ }\bibfield  {title} {\bibinfo {title} {{Stochastic Climate
  Theory and Modeling}},\ }\href@noop {} {\bibfield  {journal} {\bibinfo
  {journal} {Wiley Interdisciplinary Reviews: Climate Change}\ }\textbf
  {\bibinfo {volume} {6}},\ \bibinfo {pages} {63} (\bibinfo {year}
  {2015})}\BibitemShut {NoStop}%
\bibitem [{\citenamefont {Gunawardena}(2014)}]{Gunawardena2013}%
  \BibitemOpen
  \bibfield  {author} {\bibinfo {author} {\bibfnamefont {J.}~\bibnamefont
  {Gunawardena}},\ }\bibfield  {title} {\bibinfo {title} {{Time-scale
  separation – Michaelis and Menten’s old idea, still bearing fruit}},\
  }\href@noop {} {\bibfield  {journal} {\bibinfo  {journal} {FEBS Journal}\
  }\textbf {\bibinfo {volume} {281}},\ \bibinfo {pages} {473 } (\bibinfo {year}
  {2014})}\BibitemShut {NoStop}%
\bibitem [{\citenamefont {Tepper}\ \emph {et~al.}(2024)\citenamefont {Tepper},
  \citenamefont {Dalton},\ and\ \citenamefont {Netz}}]{Tepper2024}%
  \BibitemOpen
  \bibfield  {author} {\bibinfo {author} {\bibfnamefont {L.}~\bibnamefont
  {Tepper}}, \bibinfo {author} {\bibfnamefont {B.~A.}\ \bibnamefont {Dalton}},\
  and\ \bibinfo {author} {\bibfnamefont {R.~R.}\ \bibnamefont {Netz}},\
  }\bibfield  {title} {\bibinfo {title} {{Accurate Memory Kernel Extraction
  from Discretized Time-Series Data}},\ }\href@noop {} {\bibfield  {journal}
  {\bibinfo  {journal} {Journal of Chemical Theory and Computation}\ }\textbf
  {\bibinfo {volume} {20}},\ \bibinfo {pages} {3061} (\bibinfo {year}
  {2024})}\BibitemShut {NoStop}%
\bibitem [{\citenamefont {Changbon}\ and\ \citenamefont
  {Thirumalai}(2011)}]{Thirumalai2011}%
  \BibitemOpen
  \bibfield  {author} {\bibinfo {author} {\bibfnamefont {H.}~\bibnamefont
  {Changbon}}\ and\ \bibinfo {author} {\bibfnamefont {D.}~\bibnamefont
  {Thirumalai}},\ }\bibfield  {title} {\bibinfo {title} {{Capturing the essence
  of folding and functions of biomolecules using coarse-grained models}},\
  }\href@noop {} {\bibfield  {journal} {\bibinfo  {journal} {Nature
  Communications}\ }\textbf {\bibinfo {volume} {2}},\ \bibinfo {pages} {487}
  (\bibinfo {year} {2011})}\BibitemShut {NoStop}%
\bibitem [{\citenamefont {Denesyuk}\ and\ \citenamefont
  {Thirumalai}(2013)}]{Thirumalai2013}%
  \BibitemOpen
  \bibfield  {author} {\bibinfo {author} {\bibfnamefont {N.~A.}\ \bibnamefont
  {Denesyuk}}\ and\ \bibinfo {author} {\bibfnamefont {D.}~\bibnamefont
  {Thirumalai}},\ }\bibfield  {title} {\bibinfo {title} {{Coarse-graining DNA
  for simulations of DNA nanotechnology}},\ }\href@noop {} {\bibfield
  {journal} {\bibinfo  {journal} {Phys. Chem. Chem. Phys.}\ }\textbf {\bibinfo
  {volume} {15}},\ \bibinfo {pages} {20395} (\bibinfo {year}
  {2013})}\BibitemShut {NoStop}%
\bibitem [{\citenamefont {Doye}(2014)}]{Doye2013}%
  \BibitemOpen
  \bibfield  {author} {\bibinfo {author} {\bibfnamefont {J.~P. K. e.~a.}\
  \bibnamefont {Doye}},\ }\bibfield  {title} {\bibinfo {title} {{Time-scale
  separation – Michaelis and Menten’s old idea, still bearing fruit}},\
  }\href@noop {} {\bibfield  {journal} {\bibinfo  {journal} {FEBS Journal}\
  }\textbf {\bibinfo {volume} {281}},\ \bibinfo {pages} {473 } (\bibinfo {year}
  {2014})}\BibitemShut {NoStop}%
\bibitem [{\citenamefont {Potoyan}\ \emph {et~al.}(2013)\citenamefont
  {Potoyan}, \citenamefont {Savelyev},\ and\ \citenamefont
  {Papoian}}]{Papoian2013}%
  \BibitemOpen
  \bibfield  {author} {\bibinfo {author} {\bibfnamefont {D.~A.}\ \bibnamefont
  {Potoyan}}, \bibinfo {author} {\bibfnamefont {A.}~\bibnamefont {Savelyev}},\
  and\ \bibinfo {author} {\bibfnamefont {G.~A.}\ \bibnamefont {Papoian}},\
  }\bibfield  {title} {\bibinfo {title} {{Recent successes in coarse-grained
  modeling of DNA }},\ }\href@noop {} {\bibfield  {journal} {\bibinfo
  {journal} {WIREs Comput Mol Sci}\ }\textbf {\bibinfo {volume} {3}},\ \bibinfo
  {pages} {69} (\bibinfo {year} {2013})}\BibitemShut {NoStop}%
\bibitem [{\citenamefont {Souza}(2021)}]{Marrink2021}%
  \BibitemOpen
  \bibfield  {author} {\bibinfo {author} {\bibfnamefont {P.~C. T. e.~a.}\
  \bibnamefont {Souza}},\ }\bibfield  {title} {\bibinfo {title} {{Martini 3: a
  general purpose force field for coarse-grained molecular dynamics}},\
  }\href@noop {} {\bibfield  {journal} {\bibinfo  {journal} {Nature Methods}\
  }\textbf {\bibinfo {volume} {18}},\ \bibinfo {pages} {382} (\bibinfo {year}
  {2021})}\BibitemShut {NoStop}%
\bibitem [{\citenamefont {Wang}\ \emph {et~al.}(2019)\citenamefont {Wang},
  \citenamefont {Olsson}, \citenamefont {Wehmeyer}, \citenamefont {Perez},
  \citenamefont {Charron}, \citenamefont {de~Fabritiis}, \citenamefont {Noe},\
  and\ \citenamefont {Clementi}}]{Clementi2019}%
  \BibitemOpen
  \bibfield  {author} {\bibinfo {author} {\bibfnamefont {J.}~\bibnamefont
  {Wang}}, \bibinfo {author} {\bibfnamefont {S.}~\bibnamefont {Olsson}},
  \bibinfo {author} {\bibfnamefont {C.}~\bibnamefont {Wehmeyer}}, \bibinfo
  {author} {\bibfnamefont {A.}~\bibnamefont {Perez}}, \bibinfo {author}
  {\bibfnamefont {N.~E.}\ \bibnamefont {Charron}}, \bibinfo {author}
  {\bibfnamefont {G.}~\bibnamefont {de~Fabritiis}}, \bibinfo {author}
  {\bibfnamefont {F.}~\bibnamefont {Noe}},\ and\ \bibinfo {author}
  {\bibfnamefont {C.}~\bibnamefont {Clementi}},\ }\bibfield  {title} {\bibinfo
  {title} {{Machine Learning of Coarse-Grained Molecular Dynamics Force
  Fields}},\ }\href@noop {} {\bibfield  {journal} {\bibinfo  {journal} {ACS
  Cent. Sci.}\ }\textbf {\bibinfo {volume} {5}},\ \bibinfo {pages} {755}
  (\bibinfo {year} {2019})}\BibitemShut {NoStop}%
\bibitem [{\citenamefont {Ayaz}\ \emph {et~al.}(2021)\citenamefont {Ayaz},
  \citenamefont {Tepper}, \citenamefont {Br{\"u}nig}, \citenamefont {Kappler},
  \citenamefont {Daldrop},\ and\ \citenamefont {Netz}}]{ayaz2021}%
  \BibitemOpen
  \bibfield  {author} {\bibinfo {author} {\bibfnamefont {C.}~\bibnamefont
  {Ayaz}}, \bibinfo {author} {\bibfnamefont {L.}~\bibnamefont {Tepper}},
  \bibinfo {author} {\bibfnamefont {F.~N.}\ \bibnamefont {Br{\"u}nig}},
  \bibinfo {author} {\bibfnamefont {J.}~\bibnamefont {Kappler}}, \bibinfo
  {author} {\bibfnamefont {J.~O.}\ \bibnamefont {Daldrop}},\ and\ \bibinfo
  {author} {\bibfnamefont {R.~R.}\ \bibnamefont {Netz}},\ }\bibfield  {title}
  {\bibinfo {title} {Non-{Markovian} modeling of protein folding},\ }\bibfield
  {journal} {\bibinfo  {journal} {Proceedings of the National Academy of
  Sciences}\ }\textbf {\bibinfo {volume} {118}},\ \href
  {https://doi.org/10.1073/pnas.2023856118} {10.1073/pnas.2023856118} (\bibinfo
  {year} {2021}),\ \bibinfo {note} {publisher: National Academy of Sciences
  Section: Physical Sciences}\BibitemShut {NoStop}%
\bibitem [{\citenamefont {Dalton}\ \emph {et~al.}(2023)\citenamefont {Dalton},
  \citenamefont {Ayaz}, \citenamefont {Kiefer}, \citenamefont {Klimek},
  \citenamefont {Tepper},\ and\ \citenamefont {Netz}}]{dalton2023}%
  \BibitemOpen
  \bibfield  {author} {\bibinfo {author} {\bibfnamefont {B.~A.}\ \bibnamefont
  {Dalton}}, \bibinfo {author} {\bibfnamefont {C.}~\bibnamefont {Ayaz}},
  \bibinfo {author} {\bibfnamefont {H.}~\bibnamefont {Kiefer}}, \bibinfo
  {author} {\bibfnamefont {A.}~\bibnamefont {Klimek}}, \bibinfo {author}
  {\bibfnamefont {L.}~\bibnamefont {Tepper}},\ and\ \bibinfo {author}
  {\bibfnamefont {R.~R.}\ \bibnamefont {Netz}},\ }\bibfield  {title} {\bibinfo
  {title} {Fast protein folding is governed by memory-dependent friction},\
  }\href@noop {} {\bibfield  {journal} {\bibinfo  {journal} {Proceedings of the
  National Academy of Sciences}\ }\textbf {\bibinfo {volume} {120}},\ \bibinfo
  {pages} {e2220068120} (\bibinfo {year} {2023})}\BibitemShut {NoStop}%
\bibitem [{\citenamefont {Schmitt}\ and\ \citenamefont
  {Schulz}(2006)}]{schmitt_analyzing_2006}%
  \BibitemOpen
  \bibfield  {author} {\bibinfo {author} {\bibfnamefont {D.~T.}\ \bibnamefont
  {Schmitt}}\ and\ \bibinfo {author} {\bibfnamefont {M.}~\bibnamefont
  {Schulz}},\ }\bibfield  {title} {\bibinfo {title} {{Analyzing Memory Effects
  of Complex Systems from Time Series}},\ }\href
  {https://doi.org/10.1103/PhysRevE.73.056204} {\bibfield  {journal} {\bibinfo
  {journal} {Physical Review E}\ }\textbf {\bibinfo {volume} {73}},\ \bibinfo
  {pages} {056204} (\bibinfo {year} {2006})}\BibitemShut {NoStop}%
\bibitem [{\citenamefont {Prigogine}\ and\ \citenamefont
  {Mazur}(1953)}]{Mazur1953}%
  \BibitemOpen
  \bibfield  {author} {\bibinfo {author} {\bibfnamefont {I.}~\bibnamefont
  {Prigogine}}\ and\ \bibinfo {author} {\bibfnamefont {P.}~\bibnamefont
  {Mazur}},\ }\bibfield  {title} {\bibinfo {title} {Sur l'extension de la
  thermodynamique aux phenomenes irreversibles lies aux degres de liberte
  internes},\ }\href@noop {} {\bibfield  {journal} {\bibinfo  {journal}
  {Physica}\ }\textbf {\bibinfo {volume} {XIX}},\ \bibinfo {pages} {241}
  (\bibinfo {year} {1953})}\BibitemShut {NoStop}%
\bibitem [{\citenamefont {Lebowitz}(1959)}]{Lebowitz1959}%
  \BibitemOpen
  \bibfield  {author} {\bibinfo {author} {\bibfnamefont {J.~L.}\ \bibnamefont
  {Lebowitz}},\ }\bibfield  {title} {\bibinfo {title} {Stationary
  nonequilibrium {Gibbsian} ensembles},\ }\href@noop {} {\bibfield  {journal}
  {\bibinfo  {journal} {Phys. Rev.}\ }\textbf {\bibinfo {volume} {114}},\
  \bibinfo {pages} {1192} (\bibinfo {year} {1959})}\BibitemShut {NoStop}%
\bibitem [{\citenamefont {Zwanzig}(1960)}]{zwanzig_ensemble_1960}%
  \BibitemOpen
  \bibfield  {author} {\bibinfo {author} {\bibfnamefont {R.}~\bibnamefont
  {Zwanzig}},\ }\bibfield  {title} {\bibinfo {title} {Ensemble method in the
  theory of irreversibility},\ }\href {https://doi.org/10.1063/1.1731409}
  {\bibfield  {journal} {\bibinfo  {journal} {The Journal of Chemical Physics}\
  }\textbf {\bibinfo {volume} {33}},\ \bibinfo {pages} {1338} (\bibinfo {year}
  {1960})}\BibitemShut {NoStop}%
\bibitem [{\citenamefont {de~Groot}\ and\ \citenamefont
  {Mazur}(1962)}]{deGroot}%
  \BibitemOpen
  \bibfield  {author} {\bibinfo {author} {\bibfnamefont {S.~R.}\ \bibnamefont
  {de~Groot}}\ and\ \bibinfo {author} {\bibfnamefont {P.}~\bibnamefont
  {Mazur}},\ }\href@noop {} {\emph {\bibinfo {title} {Non-Equilibrium
  Thermodynamics}}}\ (\bibinfo  {publisher} {North-Holland Pub. Co.,
  Amsterdam},\ \bibinfo {year} {1962})\BibitemShut {NoStop}%
\bibitem [{\citenamefont {Schmittmann}\ and\ \citenamefont {Zia}(1998)}]{Zia}%
  \BibitemOpen
  \bibfield  {author} {\bibinfo {author} {\bibfnamefont {B.}~\bibnamefont
  {Schmittmann}}\ and\ \bibinfo {author} {\bibfnamefont {R.~K.~P.}\
  \bibnamefont {Zia}},\ }\bibfield  {title} {\bibinfo {title} {Driven diffusive
  systems. an introduction and recent developments},\ }\href@noop {} {\bibfield
   {journal} {\bibinfo  {journal} {Phys. Rep.}\ }\textbf {\bibinfo {volume}
  {301}},\ \bibinfo {pages} {45} (\bibinfo {year} {1998})}\BibitemShut
  {NoStop}%
\bibitem [{\citenamefont {Derrida}\ \emph {et~al.}(2001)\citenamefont
  {Derrida}, \citenamefont {Lebowitz},\ and\ \citenamefont
  {Speer}}]{Derrida2001}%
  \BibitemOpen
  \bibfield  {author} {\bibinfo {author} {\bibfnamefont {B.}~\bibnamefont
  {Derrida}}, \bibinfo {author} {\bibfnamefont {J.~L.}\ \bibnamefont
  {Lebowitz}},\ and\ \bibinfo {author} {\bibfnamefont {E.~R.}\ \bibnamefont
  {Speer}},\ }\bibfield  {title} {\bibinfo {title} {Free energy functional for
  nonequilibrium systems: An exactly solvable case},\ }\href@noop {} {\bibfield
   {journal} {\bibinfo  {journal} {Phys. Rev. Lett.}\ }\textbf {\bibinfo
  {volume} {87}},\ \bibinfo {pages} {150601} (\bibinfo {year}
  {2001})}\BibitemShut {NoStop}%
\bibitem [{\citenamefont {Ilg}\ and\ \citenamefont
  {Barrat}(2007)}]{Barrat2007}%
  \BibitemOpen
  \bibfield  {author} {\bibinfo {author} {\bibfnamefont {P.}~\bibnamefont
  {Ilg}}\ and\ \bibinfo {author} {\bibfnamefont {J.~L.}\ \bibnamefont
  {Barrat}},\ }\bibfield  {title} {\bibinfo {title} {From single-particle to
  collective effective temperatures in an active fluid of self-propelled
  particles},\ }\href@noop {} {\bibfield  {journal} {\bibinfo  {journal}
  {Europhys. Lett.}\ }\textbf {\bibinfo {volume} {111}},\ \bibinfo {pages}
  {26001} (\bibinfo {year} {2007})}\BibitemShut {NoStop}%
\bibitem [{\citenamefont {Fodor}\ \emph {et~al.}(2016)\citenamefont {Fodor},
  \citenamefont {Nardini}, \citenamefont {Cates}, \citenamefont {Tailleur},
  \citenamefont {Visco},\ and\ \citenamefont {van Wijland}}]{Fodor2016}%
  \BibitemOpen
  \bibfield  {author} {\bibinfo {author} {\bibfnamefont {E.}~\bibnamefont
  {Fodor}}, \bibinfo {author} {\bibfnamefont {C.}~\bibnamefont {Nardini}},
  \bibinfo {author} {\bibfnamefont {M.~E.}\ \bibnamefont {Cates}}, \bibinfo
  {author} {\bibfnamefont {J.}~\bibnamefont {Tailleur}}, \bibinfo {author}
  {\bibfnamefont {P.}~\bibnamefont {Visco}},\ and\ \bibinfo {author}
  {\bibfnamefont {F.}~\bibnamefont {van Wijland}},\ }\bibfield  {title}
  {\bibinfo {title} {How far from equilibrium is active matter?},\ }\href@noop
  {} {\bibfield  {journal} {\bibinfo  {journal} {Phys. Rev. Lett.}\ }\textbf
  {\bibinfo {volume} {117}},\ \bibinfo {pages} {038103} (\bibinfo {year}
  {2016})}\BibitemShut {NoStop}%
\bibitem [{\citenamefont {Jarzynski}(2000)}]{Jarzynski2000}%
  \BibitemOpen
  \bibfield  {author} {\bibinfo {author} {\bibfnamefont {C.}~\bibnamefont
  {Jarzynski}},\ }\bibfield  {title} {\bibinfo {title} {Hamiltonian derivation
  of a detailed fluctuation theorem},\ }\href@noop {} {\bibfield  {journal}
  {\bibinfo  {journal} {Journal of Statistical Physics}\ }\textbf {\bibinfo
  {volume} {98}},\ \bibinfo {pages} {77} (\bibinfo {year} {2000})}\BibitemShut
  {NoStop}%
\bibitem [{\citenamefont {Harada}\ and\ \citenamefont
  {Sasa}(2005)}]{Harada2005}%
  \BibitemOpen
  \bibfield  {author} {\bibinfo {author} {\bibfnamefont {T.}~\bibnamefont
  {Harada}}\ and\ \bibinfo {author} {\bibfnamefont {S.~I.}\ \bibnamefont
  {Sasa}},\ }\bibfield  {title} {\bibinfo {title} {Equality connecting energy
  dissipation with a violation of the fluctuation-response relation},\
  }\href@noop {} {\bibfield  {journal} {\bibinfo  {journal} {Phys. Rev. Lett.}\
  }\textbf {\bibinfo {volume} {95}},\ \bibinfo {pages} {130602} (\bibinfo
  {year} {2005})}\BibitemShut {NoStop}%
\bibitem [{\citenamefont {Seifert}(2005)}]{Seifert2005}%
  \BibitemOpen
  \bibfield  {author} {\bibinfo {author} {\bibfnamefont {U.}~\bibnamefont
  {Seifert}},\ }\bibfield  {title} {\bibinfo {title} {Entropy production along
  a stochastic trajectory and an integral fluctuation theorem},\ }\href@noop {}
  {\bibfield  {journal} {\bibinfo  {journal} {Phys. Rev. Lett.}\ }\textbf
  {\bibinfo {volume} {95}},\ \bibinfo {pages} {040602} (\bibinfo {year}
  {2005})}\BibitemShut {NoStop}%
\bibitem [{\citenamefont {Prost}\ \emph {et~al.}(2009)\citenamefont {Prost},
  \citenamefont {Joanny},\ and\ \citenamefont {Parrondo}}]{Prost2009}%
  \BibitemOpen
  \bibfield  {author} {\bibinfo {author} {\bibfnamefont {J.}~\bibnamefont
  {Prost}}, \bibinfo {author} {\bibfnamefont {J.-F.}\ \bibnamefont {Joanny}},\
  and\ \bibinfo {author} {\bibfnamefont {J.~M.~R.}\ \bibnamefont {Parrondo}},\
  }\bibfield  {title} {\bibinfo {title} {Generalized fluctuation-dissipation
  theorem for steady-state systems},\ }\href@noop {} {\bibfield  {journal}
  {\bibinfo  {journal} {Phys. Rev. Lett.}\ }\textbf {\bibinfo {volume} {103}},\
  \bibinfo {pages} {090601} (\bibinfo {year} {2009})}\BibitemShut {NoStop}%
\bibitem [{\citenamefont {Baiesi}\ \emph {et~al.}(2009)\citenamefont {Baiesi},
  \citenamefont {Maes},\ and\ \citenamefont {Wynants}}]{Wynants2009}%
  \BibitemOpen
  \bibfield  {author} {\bibinfo {author} {\bibfnamefont {M.}~\bibnamefont
  {Baiesi}}, \bibinfo {author} {\bibfnamefont {C.}~\bibnamefont {Maes}},\ and\
  \bibinfo {author} {\bibfnamefont {B.}~\bibnamefont {Wynants}},\ }\bibfield
  {title} {\bibinfo {title} {Fluctuations and response of nonequilibrium
  states},\ }\href@noop {} {\bibfield  {journal} {\bibinfo  {journal} {Phys.
  Rev. Lett.}\ }\textbf {\bibinfo {volume} {103}},\ \bibinfo {pages} {010602}
  (\bibinfo {year} {2009})}\BibitemShut {NoStop}%
\bibitem [{\citenamefont {Seifert}\ and\ \citenamefont
  {Speck}(2010)}]{Seifert2010}%
  \BibitemOpen
  \bibfield  {author} {\bibinfo {author} {\bibfnamefont {U.}~\bibnamefont
  {Seifert}}\ and\ \bibinfo {author} {\bibfnamefont {T.}~\bibnamefont
  {Speck}},\ }\bibfield  {title} {\bibinfo {title} {Fluctuation-dissipation
  theorem in nonequilibrium steady states},\ }\href@noop {} {\bibfield
  {journal} {\bibinfo  {journal} {Europhys. Lett.}\ }\textbf {\bibinfo {volume}
  {89}},\ \bibinfo {pages} {10007} (\bibinfo {year} {2010})}\BibitemShut
  {NoStop}%
\bibitem [{\citenamefont {Netz}(2020)}]{Netz2020}%
  \BibitemOpen
  \bibfield  {author} {\bibinfo {author} {\bibfnamefont {R.~R.}\ \bibnamefont
  {Netz}},\ }\bibfield  {title} {\bibinfo {title} {Approach to equilibrium and
  nonequilibrium stationary distributions of interacting many-particle systems
  that are coupled to different heat baths},\ }\href@noop {} {\bibfield
  {journal} {\bibinfo  {journal} {Phys. Rev. E}\ }\textbf {\bibinfo {volume}
  {101}},\ \bibinfo {pages} {022120} (\bibinfo {year} {2020})}\BibitemShut
  {NoStop}%
\bibitem [{\citenamefont {Mori}(1965)}]{mori_transport_1965}%
  \BibitemOpen
  \bibfield  {author} {\bibinfo {author} {\bibfnamefont {H.}~\bibnamefont
  {Mori}},\ }\bibfield  {title} {\bibinfo {title} {Transport, {Collective}
  {Motion}, and {Brownian} {Motion}},\ }\href
  {https://doi.org/10.1143/PTP.33.423} {\bibfield  {journal} {\bibinfo
  {journal} {Progress of Theoretical Physics}\ }\textbf {\bibinfo {volume}
  {33}},\ \bibinfo {pages} {423} (\bibinfo {year} {1965})}\BibitemShut
  {NoStop}%
\bibitem [{\citenamefont {Zwanzig}(1961)}]{zwanzig1961memory}%
  \BibitemOpen
  \bibfield  {author} {\bibinfo {author} {\bibfnamefont {R.}~\bibnamefont
  {Zwanzig}},\ }\bibfield  {title} {\bibinfo {title} {Memory {Effects} in
  {Irreversible} {Thermodynamics}},\ }\href
  {https://doi.org/10.1103/PhysRev.124.983} {\bibfield  {journal} {\bibinfo
  {journal} {Physical Review}\ }\textbf {\bibinfo {volume} {124}},\ \bibinfo
  {pages} {983} (\bibinfo {year} {1961})}\BibitemShut {NoStop}%
\bibitem [{\citenamefont {Straub}\ \emph {et~al.}(1987)\citenamefont {Straub},
  \citenamefont {Borkovec},\ and\ \citenamefont {Berne}}]{straub1987}%
  \BibitemOpen
  \bibfield  {author} {\bibinfo {author} {\bibfnamefont {J.~E.}\ \bibnamefont
  {Straub}}, \bibinfo {author} {\bibfnamefont {M.}~\bibnamefont {Borkovec}},\
  and\ \bibinfo {author} {\bibfnamefont {B.~J.}\ \bibnamefont {Berne}},\
  }\bibfield  {title} {\bibinfo {title} {{Calculation of dynamic friction on
  intramolecular degrees of freedom}},\ }\href@noop {} {\bibfield  {journal}
  {\bibinfo  {journal} {The Journal of Physical Chemistry}\ }\textbf {\bibinfo
  {volume} {91}},\ \bibinfo {pages} {4995} (\bibinfo {year}
  {1987})}\BibitemShut {NoStop}%
\bibitem [{\citenamefont {Daldrop}\ \emph {et~al.}(2018)\citenamefont
  {Daldrop}, \citenamefont {Kappler}, \citenamefont {Br{\"u}nig},\ and\
  \citenamefont {Netz}}]{daldrop_butane_2018}%
  \BibitemOpen
  \bibfield  {author} {\bibinfo {author} {\bibfnamefont {J.~O.}\ \bibnamefont
  {Daldrop}}, \bibinfo {author} {\bibfnamefont {J.}~\bibnamefont {Kappler}},
  \bibinfo {author} {\bibfnamefont {F.~N.}\ \bibnamefont {Br{\"u}nig}},\ and\
  \bibinfo {author} {\bibfnamefont {R.~R.}\ \bibnamefont {Netz}},\ }\bibfield
  {title} {\bibinfo {title} {Butane dihedral angle dynamics in water is
  dominated by internal friction},\ }\href
  {https://doi.org/10.1073/pnas.1722327115} {\bibfield  {journal} {\bibinfo
  {journal} {Proceedings of the National Academy of Sciences}\ }\textbf
  {\bibinfo {volume} {115}},\ \bibinfo {pages} {5169} (\bibinfo {year}
  {2018})},\ \bibinfo {note} {publisher: National Academy of Sciences Section:
  Biological Sciences}\BibitemShut {NoStop}%
\bibitem [{\citenamefont {Kowalik}\ \emph {et~al.}(2019)\citenamefont
  {Kowalik}, \citenamefont {Daldrop}, \citenamefont {Kappler}, \citenamefont
  {Schulz}, \citenamefont {Schlaich},\ and\ \citenamefont
  {Netz}}]{kowalik2019}%
  \BibitemOpen
  \bibfield  {author} {\bibinfo {author} {\bibfnamefont {B.}~\bibnamefont
  {Kowalik}}, \bibinfo {author} {\bibfnamefont {J.~O.}\ \bibnamefont
  {Daldrop}}, \bibinfo {author} {\bibfnamefont {J.}~\bibnamefont {Kappler}},
  \bibinfo {author} {\bibfnamefont {J.~C.~F.}\ \bibnamefont {Schulz}}, \bibinfo
  {author} {\bibfnamefont {A.}~\bibnamefont {Schlaich}},\ and\ \bibinfo
  {author} {\bibfnamefont {R.~R.}\ \bibnamefont {Netz}},\ }\bibfield  {title}
  {\bibinfo {title} {{Memory-kernel extraction for different molecular solutes
  in solvents of varying viscosity in confinement}},\ }\href@noop {} {\bibfield
   {journal} {\bibinfo  {journal} {Physical Review E}\ }\textbf {\bibinfo
  {volume} {100}},\ \bibinfo {pages} {012126} (\bibinfo {year}
  {2019})}\BibitemShut {NoStop}%
\bibitem [{\citenamefont {Grogan}\ \emph {et~al.}(2020)\citenamefont {Grogan},
  \citenamefont {Lei}, \citenamefont {Li},\ and\ \citenamefont
  {Baker}}]{grogan_data-driven_2020}%
  \BibitemOpen
  \bibfield  {author} {\bibinfo {author} {\bibfnamefont {F.}~\bibnamefont
  {Grogan}}, \bibinfo {author} {\bibfnamefont {H.}~\bibnamefont {Lei}},
  \bibinfo {author} {\bibfnamefont {X.}~\bibnamefont {Li}},\ and\ \bibinfo
  {author} {\bibfnamefont {N.~A.}\ \bibnamefont {Baker}},\ }\bibfield  {title}
  {\bibinfo {title} {{Data-Driven Molecular Modeling with the Generalized
  Langevin Equation}},\ }\href@noop {} {\bibfield  {journal} {\bibinfo
  {journal} {Journal of Computational Physics}\ }\textbf {\bibinfo {volume}
  {418}},\ \bibinfo {pages} {109633} (\bibinfo {year} {2020})}\BibitemShut
  {NoStop}%
\bibitem [{\citenamefont {Satija}\ and\ \citenamefont
  {Makarov}(2019)}]{satija2019}%
  \BibitemOpen
  \bibfield  {author} {\bibinfo {author} {\bibfnamefont {R.}~\bibnamefont
  {Satija}}\ and\ \bibinfo {author} {\bibfnamefont {D.~E.}\ \bibnamefont
  {Makarov}},\ }\bibfield  {title} {\bibinfo {title} {Generalized {Langevin}
  {Equation} as a {Model} for {Barrier} {Crossing} {Dynamics} in {Biomolecular}
  {Folding}},\ }\href {https://doi.org/10.1021/acs.jpcb.8b11137} {\bibfield
  {journal} {\bibinfo  {journal} {The Journal of Physical Chemistry B}\
  }\textbf {\bibinfo {volume} {123}},\ \bibinfo {pages} {802} (\bibinfo {year}
  {2019})},\ \bibinfo {note} {publisher: American Chemical Society}\BibitemShut
  {NoStop}%
\bibitem [{\citenamefont {Br{\"{u}}nig}\ \emph
  {et~al.}(2022{\natexlab{a}})\citenamefont {Br{\"{u}}nig}, \citenamefont
  {Geburtig}, \citenamefont {von Canal}, \citenamefont {Kappler},\ and\
  \citenamefont {Netz}}]{brunig2022a}%
  \BibitemOpen
  \bibfield  {author} {\bibinfo {author} {\bibfnamefont {F.~N.}\ \bibnamefont
  {Br{\"{u}}nig}}, \bibinfo {author} {\bibfnamefont {O.}~\bibnamefont
  {Geburtig}}, \bibinfo {author} {\bibfnamefont {A.}~\bibnamefont {von Canal}},
  \bibinfo {author} {\bibfnamefont {J.}~\bibnamefont {Kappler}},\ and\ \bibinfo
  {author} {\bibfnamefont {R.~R.}\ \bibnamefont {Netz}},\ }\bibfield  {title}
  {\bibinfo {title} {{Time-Dependent Friction Effects on Vibrational Infrared
  Frequencies and Line Shapes of Liquid Water}},\ }\href@noop {} {\bibfield
  {journal} {\bibinfo  {journal} {The Journal of Physical Chemistry B}\
  }\textbf {\bibinfo {volume} {126}},\ \bibinfo {pages} {1579} (\bibinfo {year}
  {2022}{\natexlab{a}})}\BibitemShut {NoStop}%
\bibitem [{\citenamefont {Br{\"{u}}nig}\ \emph
  {et~al.}(2022{\natexlab{b}})\citenamefont {Br{\"{u}}nig}, \citenamefont
  {Daldrop},\ and\ \citenamefont {Netz}}]{brunig2022d}%
  \BibitemOpen
  \bibfield  {author} {\bibinfo {author} {\bibfnamefont {F.~N.}\ \bibnamefont
  {Br{\"{u}}nig}}, \bibinfo {author} {\bibfnamefont {J.~O.}\ \bibnamefont
  {Daldrop}},\ and\ \bibinfo {author} {\bibfnamefont {R.~R.}\ \bibnamefont
  {Netz}},\ }\bibfield  {title} {\bibinfo {title} {{Pair-Reaction Dynamics in
  Water: Competition of Memory, Potential Shape, and Inertial Effects}},\
  }\href@noop {} {\bibfield  {journal} {\bibinfo  {journal} {The Journal of
  Physical Chemistry B}\ }\textbf {\bibinfo {volume} {126}},\ \bibinfo {pages}
  {10295} (\bibinfo {year} {2022}{\natexlab{b}})}\BibitemShut {NoStop}%
\bibitem [{\citenamefont {Dalton}\ \emph {et~al.}(2024)\citenamefont {Dalton},
  \citenamefont {Kiefer},\ and\ \citenamefont {Netz}}]{dalton2024}%
  \BibitemOpen
  \bibfield  {author} {\bibinfo {author} {\bibfnamefont {B.~A.}\ \bibnamefont
  {Dalton}}, \bibinfo {author} {\bibfnamefont {H.}~\bibnamefont {Kiefer}},\
  and\ \bibinfo {author} {\bibfnamefont {R.~R.}\ \bibnamefont {Netz}},\
  }\bibfield  {title} {\bibinfo {title} {{The Role of Memory-Dependent Friction
  and Solvent Viscosity in Isomerization Kinetics in Viscogenic Media}},\
  }\href@noop {} {\bibfield  {journal} {\bibinfo  {journal} {Nature
  Communications}\ }\textbf {\bibinfo {volume} {15}},\ \bibinfo {pages} {3761}
  (\bibinfo {year} {2024})}\BibitemShut {NoStop}%
\bibitem [{\citenamefont {Mitterwallner}\ \emph {et~al.}(2020)\citenamefont
  {Mitterwallner}, \citenamefont {Schreiber}, \citenamefont {Daldrop},
  \citenamefont {Rädler},\ and\ \citenamefont {Netz}}]{mitterwallner2020}%
  \BibitemOpen
  \bibfield  {author} {\bibinfo {author} {\bibfnamefont {B.~G.}\ \bibnamefont
  {Mitterwallner}}, \bibinfo {author} {\bibfnamefont {C.}~\bibnamefont
  {Schreiber}}, \bibinfo {author} {\bibfnamefont {J.~O.}\ \bibnamefont
  {Daldrop}}, \bibinfo {author} {\bibfnamefont {J.~O.}\ \bibnamefont
  {Rädler}},\ and\ \bibinfo {author} {\bibfnamefont {R.~R.}\ \bibnamefont
  {Netz}},\ }\bibfield  {title} {\bibinfo {title} {{Non-Markovian data-driven
  modeling of single-cell motility}},\ }\href@noop {} {\bibfield  {journal}
  {\bibinfo  {journal} {Physical Review E}\ }\textbf {\bibinfo {volume}
  {101}},\ \bibinfo {pages} {032408} (\bibinfo {year} {2020})}\BibitemShut
  {NoStop}%
\bibitem [{\citenamefont {Hassanibesheli}\ \emph {et~al.}(2020)\citenamefont
  {Hassanibesheli}, \citenamefont {Boers},\ and\ \citenamefont
  {Kurths}}]{hassanibesheli2020reconstructing}%
  \BibitemOpen
  \bibfield  {author} {\bibinfo {author} {\bibfnamefont {F.}~\bibnamefont
  {Hassanibesheli}}, \bibinfo {author} {\bibfnamefont {N.}~\bibnamefont
  {Boers}},\ and\ \bibinfo {author} {\bibfnamefont {J.}~\bibnamefont
  {Kurths}},\ }\bibfield  {title} {\bibinfo {title} {{Reconstructing Complex
  System Dynamics from Time Series: A Method Comparison}},\ }\href@noop {}
  {\bibfield  {journal} {\bibinfo  {journal} {New Journal of Physics}\ }\textbf
  {\bibinfo {volume} {22}},\ \bibinfo {pages} {073053} (\bibinfo {year}
  {2020})}\BibitemShut {NoStop}%
\bibitem [{\citenamefont {Klimek}\ \emph {et~al.}(2024)\citenamefont {Klimek},
  \citenamefont {Mondal}, \citenamefont {Block}, \citenamefont {Sharma},\ and\
  \citenamefont {Netz}}]{klimek2024}%
  \BibitemOpen
  \bibfield  {author} {\bibinfo {author} {\bibfnamefont {A.}~\bibnamefont
  {Klimek}}, \bibinfo {author} {\bibfnamefont {D.}~\bibnamefont {Mondal}},
  \bibinfo {author} {\bibfnamefont {S.}~\bibnamefont {Block}}, \bibinfo
  {author} {\bibfnamefont {P.}~\bibnamefont {Sharma}},\ and\ \bibinfo {author}
  {\bibfnamefont {R.~R.}\ \bibnamefont {Netz}},\ }\bibfield  {title} {\bibinfo
  {title} {{Data-driven classification of individual cells by their
  non-Markovian motion}},\ }\href@noop {} {\bibfield  {journal} {\bibinfo
  {journal} {Biophysical Journal}\ }\textbf {\bibinfo {volume} {123}},\
  \bibinfo {pages} {1} (\bibinfo {year} {2024})}\BibitemShut {NoStop}%
\bibitem [{\citenamefont {Zwanzig}(2001)}]{zwanzig_nonequilibrium_2001}%
  \BibitemOpen
  \bibfield  {author} {\bibinfo {author} {\bibfnamefont {R.}~\bibnamefont
  {Zwanzig}},\ }\href@noop {} {\emph {\bibinfo {title} {Nonequilibrium
  statistical mechanics}}}\ (\bibinfo  {publisher} {Oxford UnivPress},\
  \bibinfo {address} {Oxford [u.a.]},\ \bibinfo {year} {2001})\BibitemShut
  {NoStop}%
\bibitem [{\citenamefont {Netz}(2024)}]{Netz2024}%
  \BibitemOpen
  \bibfield  {author} {\bibinfo {author} {\bibfnamefont {R.~R.}\ \bibnamefont
  {Netz}},\ }\bibfield  {title} {\bibinfo {title} {Derivation of the
  nonequilibrium generalized langevin equation from a time-dependent many-body
  hamiltonian},\ }\href@noop {} {\bibfield  {journal} {\bibinfo  {journal}
  {Phys. Rev. E}\ }\textbf {\bibinfo {volume} {110}},\ \bibinfo {pages}
  {014123} (\bibinfo {year} {2024})}\BibitemShut {NoStop}%
\bibitem [{\citenamefont {Dyson}(1949)}]{dyson_radiation_1949}%
  \BibitemOpen
  \bibfield  {author} {\bibinfo {author} {\bibfnamefont {F.~J.}\ \bibnamefont
  {Dyson}},\ }\bibfield  {title} {\bibinfo {title} {The {Radiation} {Theories}
  of {Tomonaga}, {Schwinger}, and {Feynman}},\ }\href
  {https://doi.org/10.1103/PhysRev.75.486} {\bibfield  {journal} {\bibinfo
  {journal} {Physical Review}\ }\textbf {\bibinfo {volume} {75}},\ \bibinfo
  {pages} {486} (\bibinfo {year} {1949})},\ \bibinfo {note} {publisher:
  American Physical Society}\BibitemShut {NoStop}%
\bibitem [{\citenamefont {Feynman}(1951)}]{feynman_operator_1951}%
  \BibitemOpen
  \bibfield  {author} {\bibinfo {author} {\bibfnamefont {R.~P.}\ \bibnamefont
  {Feynman}},\ }\bibfield  {title} {\bibinfo {title} {An {Operator} {Calculus}
  {Having} {Applications} in {Quantum} {Electrodynamics}},\ }\href
  {https://doi.org/10.1103/PhysRev.84.108} {\bibfield  {journal} {\bibinfo
  {journal} {Physical Review}\ }\textbf {\bibinfo {volume} {84}},\ \bibinfo
  {pages} {108} (\bibinfo {year} {1951})},\ \bibinfo {note} {publisher:
  American Physical Society}\BibitemShut {NoStop}%
\bibitem [{\citenamefont {Evans}(2008)}]{evans_statistical_2008}%
  \BibitemOpen
  \bibfield  {author} {\bibinfo {author} {\bibfnamefont {D.~J.}\ \bibnamefont
  {Evans}},\ }\href@noop {} {\emph {\bibinfo {title} {Statistical mechanics of
  nonequilibrium liquids / {Denis} {J}. {Evans}, {Gary} {Morriss}. [electronic
  resource]}}},\ \bibinfo {edition} {second edition.}\ ed.\ (\bibinfo
  {publisher} {University Press},\ \bibinfo {address} {Cambridge},\ \bibinfo
  {year} {2008})\ \bibinfo {note} {book Title: Statistical mechanics of
  nonequilibrium liquids}\BibitemShut {NoStop}%
\bibitem [{\citenamefont {Ayaz}\ \emph
  {et~al.}(2022{\natexlab{a}})\citenamefont {Ayaz}, \citenamefont {Scalfi},
  \citenamefont {Dalton},\ and\ \citenamefont {Netz}}]{Ayaz2022}%
  \BibitemOpen
  \bibfield  {author} {\bibinfo {author} {\bibfnamefont {C.}~\bibnamefont
  {Ayaz}}, \bibinfo {author} {\bibfnamefont {L.}~\bibnamefont {Scalfi}},
  \bibinfo {author} {\bibfnamefont {B.~A.}\ \bibnamefont {Dalton}},\ and\
  \bibinfo {author} {\bibfnamefont {R.~R.}\ \bibnamefont {Netz}},\ }\bibfield
  {title} {\bibinfo {title} {Generalized langevin equation with a nonlinear
  potential of mean force and nonlinear memory friction from a hybrid
  projection scheme},\ }\href@noop {} {\bibfield  {journal} {\bibinfo
  {journal} {Physical Review E}\ }\textbf {\bibinfo {volume} {105}},\ \bibinfo
  {pages} {054138} (\bibinfo {year} {2022}{\natexlab{a}})}\BibitemShut
  {NoStop}%
\bibitem [{\citenamefont {Vroylandt}(2022)}]{vroylandt_epl_2022}%
  \BibitemOpen
  \bibfield  {author} {\bibinfo {author} {\bibfnamefont {H.}~\bibnamefont
  {Vroylandt}},\ }\bibfield  {title} {\bibinfo {title} {On the derivation of
  the generalized langevin equation and the fluctuation-dissipation theorem},\
  }\href@noop {} {\bibfield  {journal} {\bibinfo  {journal} {Europhys. Lett.}\
  }\textbf {\bibinfo {volume} {140}},\ \bibinfo {pages} {62003} (\bibinfo
  {year} {2022})}\BibitemShut {NoStop}%
\bibitem [{\citenamefont {Ayaz}\ \emph
  {et~al.}(2022{\natexlab{b}})\citenamefont {Ayaz}, \citenamefont {Tepper},\
  and\ \citenamefont {Netz}}]{Ayaz2022b}%
  \BibitemOpen
  \bibfield  {author} {\bibinfo {author} {\bibfnamefont {C.}~\bibnamefont
  {Ayaz}}, \bibinfo {author} {\bibfnamefont {L.}~\bibnamefont {Tepper}},\ and\
  \bibinfo {author} {\bibfnamefont {R.~R.}\ \bibnamefont {Netz}},\ }\bibfield
  {title} {\bibinfo {title} {Markovian embedding of generalized langevin
  equations with a nonlinear friction kernel and configuration-dependent
  mass},\ }\href@noop {} {\bibfield  {journal} {\bibinfo  {journal} {Turk. J.
  Phys.}\ }\textbf {\bibinfo {volume} {46}},\ \bibinfo {pages} {194} (\bibinfo
  {year} {2022}{\natexlab{b}})}\BibitemShut {NoStop}%
\bibitem [{\citenamefont {Carof}\ \emph {et~al.}(2014)\citenamefont {Carof},
  \citenamefont {Vuilleumier},\ and\ \citenamefont
  {Rotenberg}}]{carof_two_2014}%
  \BibitemOpen
  \bibfield  {author} {\bibinfo {author} {\bibfnamefont {A.}~\bibnamefont
  {Carof}}, \bibinfo {author} {\bibfnamefont {R.}~\bibnamefont {Vuilleumier}},\
  and\ \bibinfo {author} {\bibfnamefont {B.}~\bibnamefont {Rotenberg}},\
  }\bibfield  {title} {\bibinfo {title} {Two algorithms to compute projected
  correlation functions in molecular dynamics simulations},\ }\href
  {https://doi.org/10.1063/1.4868653} {\bibfield  {journal} {\bibinfo
  {journal} {The Journal of Chemical Physics}\ }\textbf {\bibinfo {volume}
  {140}},\ \bibinfo {pages} {124103} (\bibinfo {year} {2014})},\ \bibinfo
  {note} {publisher: American Institute of Physics}\BibitemShut {NoStop}%
\bibitem [{\citenamefont {Lesnicki}\ \emph {et~al.}(2016)\citenamefont
  {Lesnicki}, \citenamefont {Vuilleumier}, \citenamefont {Carof},\ and\
  \citenamefont {Rotenberg}}]{lesnicki_molecular_2016}%
  \BibitemOpen
  \bibfield  {author} {\bibinfo {author} {\bibfnamefont {D.}~\bibnamefont
  {Lesnicki}}, \bibinfo {author} {\bibfnamefont {R.}~\bibnamefont
  {Vuilleumier}}, \bibinfo {author} {\bibfnamefont {A.}~\bibnamefont {Carof}},\
  and\ \bibinfo {author} {\bibfnamefont {B.}~\bibnamefont {Rotenberg}},\
  }\bibfield  {title} {\bibinfo {title} {Molecular {Hydrodynamics} from
  {Memory} {Kernels}},\ }\href {https://doi.org/10.1103/PhysRevLett.116.147804}
  {\bibfield  {journal} {\bibinfo  {journal} {Physical Review Letters}\
  }\textbf {\bibinfo {volume} {116}},\ \bibinfo {pages} {147804} (\bibinfo
  {year} {2016})},\ \bibinfo {note} {publisher: American Physical
  Society}\BibitemShut {NoStop}%
\bibitem [{\citenamefont {Vroylandt}\ \emph {et~al.}(2022)\citenamefont
  {Vroylandt}, \citenamefont {Gouden{\`e}ge}, \citenamefont {Monmarch{\'e}},
  \citenamefont {Pietrucci},\ and\ \citenamefont
  {Rotenberg}}]{vroylandt_likelihood_2022}%
  \BibitemOpen
  \bibfield  {author} {\bibinfo {author} {\bibfnamefont {H.}~\bibnamefont
  {Vroylandt}}, \bibinfo {author} {\bibfnamefont {L.}~\bibnamefont
  {Gouden{\`e}ge}}, \bibinfo {author} {\bibfnamefont {P.}~\bibnamefont
  {Monmarch{\'e}}}, \bibinfo {author} {\bibfnamefont {F.}~\bibnamefont
  {Pietrucci}},\ and\ \bibinfo {author} {\bibfnamefont {B.}~\bibnamefont
  {Rotenberg}},\ }\bibfield  {title} {\bibinfo {title} {Likelihood-based
  non-markovian models from molecular dynamics},\ }\href@noop {} {\bibfield
  {journal} {\bibinfo  {journal} {Proceedings of the National Academy of
  Sciences}\ }\textbf {\bibinfo {volume} {119}},\ \bibinfo {pages}
  {e2117586119} (\bibinfo {year} {2022})}\BibitemShut {NoStop}%
\bibitem [{\citenamefont {Meyer}\ \emph {et~al.}(2017)\citenamefont {Meyer},
  \citenamefont {Voigtmann},\ and\ \citenamefont
  {Schilling}}]{meyer_non-stationary_2017}%
  \BibitemOpen
  \bibfield  {author} {\bibinfo {author} {\bibfnamefont {H.}~\bibnamefont
  {Meyer}}, \bibinfo {author} {\bibfnamefont {T.}~\bibnamefont {Voigtmann}},\
  and\ \bibinfo {author} {\bibfnamefont {T.}~\bibnamefont {Schilling}},\
  }\bibfield  {title} {\bibinfo {title} {On the non-stationary generalized
  {Langevin} equation},\ }\href {https://doi.org/10.1063/1.5006980} {\bibfield
  {journal} {\bibinfo  {journal} {The Journal of Chemical Physics}\ }\textbf
  {\bibinfo {volume} {147}},\ \bibinfo {pages} {214110} (\bibinfo {year}
  {2017})},\ \bibinfo {note} {publisher: American Institute of
  Physics}\BibitemShut {NoStop}%
\bibitem [{\citenamefont {Meyer}\ \emph {et~al.}(2020)\citenamefont {Meyer},
  \citenamefont {Pelagejcev},\ and\ \citenamefont
  {Schilling}}]{meyer_non-markovian_2020}%
  \BibitemOpen
  \bibfield  {author} {\bibinfo {author} {\bibfnamefont {H.}~\bibnamefont
  {Meyer}}, \bibinfo {author} {\bibfnamefont {P.}~\bibnamefont {Pelagejcev}},\
  and\ \bibinfo {author} {\bibfnamefont {T.}~\bibnamefont {Schilling}},\
  }\bibfield  {title} {\bibinfo {title} {Non-{Markovian} out-of-equilibrium
  dynamics: {A} general numerical procedure to construct time-dependent memory
  kernels for coarse-grained observables},\ }\href
  {https://doi.org/10.1209/0295-5075/128/40001} {\bibfield  {journal} {\bibinfo
   {journal} {EPL (Europhysics Letters)}\ }\textbf {\bibinfo {volume} {128}},\
  \bibinfo {pages} {40001} (\bibinfo {year} {2020})},\ \bibinfo {note}
  {publisher: IOP Publishing}\BibitemShut {NoStop}%
\bibitem [{\citenamefont {te~Vrugt}\ and\ \citenamefont
  {Wittkowski}(2019)}]{Vrugt2019}%
  \BibitemOpen
  \bibfield  {author} {\bibinfo {author} {\bibfnamefont {M.}~\bibnamefont
  {te~Vrugt}}\ and\ \bibinfo {author} {\bibfnamefont {R.}~\bibnamefont
  {Wittkowski}},\ }\bibfield  {title} {\bibinfo {title} {Mori-zwanzig
  projection operator formalism for far-from-equilibrium systems with
  time-dependent hamiltonians},\ }\href@noop {} {\bibfield  {journal} {\bibinfo
   {journal} {Physical Review E}\ }\textbf {\bibinfo {volume} {99}},\ \bibinfo
  {pages} {062118} (\bibinfo {year} {2019})}\BibitemShut {NoStop}%
\end{thebibliography}%

\end{document}